\documentclass[superscriptaddress,showpacs,amssymb,10pt,reprint,aps,prd,longbibliography,nofootinbib,floatfix]{revtex4-1}
\usepackage{graphicx,epsfig,amssymb} 
\usepackage{amsmath,amsfonts, times}
\usepackage{bm} 
\usepackage{epstopdf}
\usepackage[linktocpage,colorlinks]{hyperref}
\usepackage[usenames]{color}     
\usepackage{natbib}
\usepackage{float}
\usepackage[utf8x]{inputenc}
\usepackage{soul}
\usepackage[dvipsnames]{xcolor}
\usepackage[normalem]{ulem}

\usepackage{graphicx}
\usepackage{tikz}
\usepackage{cleveref}
\usepackage{ulem} 
\usepackage{multirow}
\usepackage{placeins}
\usepackage{comment}

\begin{document}
\title{\large Hairy black holes via gravitational decoupling: light rings, absorption and spectral lines}

\author{Gabriel P. Ribeiro}
\email{gabriel.ribeiro@icen.ufpa.br}
\affiliation{Programa de P\'os-Gradua\c{c}\~{a}o em F\'{\i}sica, Universidade 
	Federal do Par\'a, 66075-110, Bel\'em, Par\'a, Brazil.}

\author{Renan B. Magalhães}
    \email{renan.batalha@ufma.br}
	\affiliation{Programa de Pós-graduação em Física, Universidade Federal do Maranhão, Campus Universitário do Bacanga, São Luís (MA), 65080-805, Brazil.}

\author{Lu\'is C. B. Crispino}
\email{crispino@ufpa.br}
\affiliation{Programa de P\'os-Gradua\c{c}\~{a}o em F\'{\i}sica, Universidade 
	Federal do Par\'a, 66075-110, Bel\'em, Par\'a, Brazil.}

\begin{abstract}
We investigate the absorption of massless scalar waves by three distinct hairy black hole solutions obtained through the gravitational decoupling method, considering  the weak, the strong or the dominant energy conditions. Remarkably, in certain configurations of hairy black holes associated with the fulfillment of the weak energy condition, quasibound states may appear, resulting in Breit-Wigner-like resonances in their absorption profile. These quasibound states (and consequently the spectral lines in the absorption spectrum) can be related to stable light rings in the spacetime, a structure often associated with horizonless exotic compact objects, such as wormholes. We investigate how the gravitational decoupling method introduces novel light ring structures in hairy black holes and influences the absorption spectra through its deformation parameters. Our numerical results show excellent agreement with well-known approximations.
\end{abstract}

\date{\today}

\maketitle

\section{Introduction}

Black holes~(BHs) figure among the most interesting predictions of General Relativity (GR). These objects are defined by a hypersurface of no return, called event horizon, that causally separates two regions of spacetime. Consequently, any causal path that crosses the event horizon must extend to the interior of the BH.
Although BH solutions were found shortly after the advent of GR, they were long considered unrealistic astrophysical objects, dismissed instead as mathematical artifacts of the theory. However, in 1965, Penrose published a seminal paper~\cite{Penrose1965} showing that, regardless of symmetries, BHs have a clear and realistic formation mechanism, the gravitational collapse. Nevertheless, he also showed that, in the process, singularities are unavoidable within GR.

Over the years, a wide variety of BH solutions have been discovered within GR and its extensions~\cite{Kerr1963, Newman1963, newman1965metric, herdeiro2014kerr, ayon1999new, yajima2001black, kastor1993cosmological, moffat2015black, konno2009rotating}. Yet, amid this diversity, one guiding principle seemed to persist: BHs are simple objects, described by a small number of parameters. In particular, it is conjectured that, within GR, BHs are described by only three parameters, namely, mass, charge, and angular momentum. This statement is typically supported by the so-called no-hair theorems~\cite{Ruffini1971, Bekenstein1972-1, Bekenstein1972-2, Bekenstein1972-3, Misner1973, Bekenstein1973}. However, by relaxing the theorems' underlying assumptions, numerous hairy solutions have been discovered, both within the framework of GR as well as in modified gravity theories.

One of the simplest and most fundamental examples is the spontaneous formation of scalar hair in BHs, which arises due to the non-minimal coupling of a scalar field to gravity~\cite{Herdeiro2018}. This spontaneous scalarization can be further extended by including a non-minimal coupling between the scalar field and the Gauss-Bonnet term~\cite{Doneva2018, Cunha2019}. Furthermore,  GR can be minimally coupled to Proca fields, leading to BHs with Proca hair~\cite{Santos2020}. Moreover, hairy BHs might emerge in the Einstein-Yang-Mills \cite{Mazharimousavi2008}, Einstein-Yang-Mills-Higgs \cite{vanDerBij2000}, Einstein-Skyrme theories \cite{Adam2016}, or even within the framework of string theory~\cite{Giddings1994}.

Due to the nonlinear nature of GR, finding hairy solutions is a challenging task, with the difficulty becoming even more pronounced in the context of modified gravity. To simplify this complexity, one strategy involves approximating GR's field equations to a \textit{quasi-linear} regime~\cite{Ovalle2008, Ovalle2010, Casadio2012, Ovalle2013, Ovalle2013role, Casadio2014}. This is achieved by assuming that the introduction of an additional matter field perturbs or ``deforms'' known exact solutions of GR. This behavior enables an effective linear decoupling of the matter fields, resulting in a set of equations that can be solved upon the imposition of appropriate energy conditions. Such approach is known as gravitational decoupling, referring to the resulting separation of the matter sector~\cite{Casadio2015, Ovalle2019}. 
Hence, starting from a seed solution and imposing a minimal set of conditions that a BH must satisfy, one can derive novel BHs using the gravitational decoupling framework~\cite{Ovalle2016, Ovalle2021, Avalos2023, Zhang2023, Tello2024}. These solutions usually present primary hair, parameters not associated with mass, charge, or angular momentum, which lead for instance, to distinct phenomenological signatures that can tell them apart from standard GR BHs.

A comprehensive way to probe the physics of BHs is through the analysis of test particles and fields in their vicinity. In this regard, an aspect of particular importance is how these compact objects absorb particles and fields~\cite{Sanchez1978,Crispino2007,Jung2005,Crispino2008, Crispino2009,Glampedakis2001, Dolan2008}.  In particular, the simplest wave-like probe one can consider is a scalar field, whose absorption has been extensively studied in various BH scenarios~\cite{Macedo2013,Leite2017, Benone2018, Benone2019,magalhaes2020schwarzschild,magalhaes2020absorption}. The scalar absorption properties of BHs exhibit universal characteristics across the frequency spectrum. For instance, in the low-frequency regime, the absorption cross section of massless scalar waves approaches the area of the event horizon~\cite{Das1997, Higuchi2001, Higuchi2002}, while in the high-frequency regime, where massless scalar waves behave similarly to null geodesics, the absorption cross section oscillates around the geometric capture cross section of the BH. 
Applying the Regge theory~\cite{Newton2013, Chandrasekhar1992, Andersson1994} reveals that these oscillations are governed by properties of unstable null circular geodesics, the so-called light rings, specifically the angular frequency and Lyapunov exponent of these geodesics~\cite{Decanini2011}.

We investigate the absorption of massless scalar fields by three distinct hairy BH spacetimes generated via gravitational decoupling, each satisfying a specific energy condition, namely the dominant energy condition (DEC), the strong energy condition (SEC), or the weak energy condition (WEC). Remarkably, the scattering potential for scalar perturbations in certain parameter ranges of these spacetimes exhibits wells, indicating the possible existence of quasibound states. Such states are typically associated with stable light rings~\cite{cardoso2014light}, features common in horizonless ultracompact objects like wormholes~\cite{Junior2020,magalhaes2023asymmetric} and weakly dissipative stars~\cite{Macedo2018}. As we shall see, the emergence of these quasibound states leads to distinct signatures in the absorption spectrum, for example the appearance of spectral lines, manifest as sharp absorption peaks in the absorption profile. Notably, our analysis reveals that stable light rings typically appear in configurations that violate the SEC outside the event horizon, which is consistent with the conjecture proposed in Ref.~\cite{Cvetivc2016}.  Remarkably, the hairy BHs associated with the WEC can support stable light rings without violating the WEC on or outside the event horizon. 

The content of this paper is organized as follows. In Sec.~\ref{sec:decoupling} we review the key ideas of the gravitational decoupling method and present the hairy BH solutions we investigate throughout the paper. In Sec.~\ref{sec:nullgeod}, we analyze the trajectories of massless particles and investigate the light ring structure of the hairy BH solutions. In Sec.~\ref{sec:absorption}, we review the partial-wave approach applied to the absorption of massless scalar waves. We present a selection of our numerical results in Sec.~\ref{sec:results}, where we examine how the additional parameter of these hairy solutions can influence the absorption cross section. We compute the partial and total absorption cross sections for various hairy solutions and compare them with the well-known results for the Schwarzschild solution (the seed solution of these hairy BHs). Moreover, we explore the appearance of quasibound states in hairy BHs associated with SEC and WEC. Furthermore, we discuss the impact of such trapped modes in the absorption profile. We present our final remarks in Sec. \ref{sec:remarks}. We use natural units, in which $G=c=\hbar=1$, and signature $(+,-,-,-)$.
 
\section{Setup}
\label{sec:decoupling}
\subsection{Gravitational decoupling method}
To deal with the non-linearity of GR, one might ``decouple'' matter sources in a quasi-linear fashion, somewhat similar to the usual behaviour of linear theories. This gives rise to the so-called gravitational decoupling method~\cite{Ovalle2019}. To do so, one considers the Einstein's equations
\begin{equation}
\label{eqn:gdecoup}
G_{\mu\nu} = \kappa^2 \tilde{T}_{\mu\nu},
\end{equation}
with the (total) energy momentum tensor being composed of multiple contributions, namely
\begin{equation}
\tilde{T}_{\mu\nu} = T_{\mu\nu} + \Theta_{\mu\nu},
\end{equation}
where $T_{\mu\nu}$ is associated with a known matter distribution that serves as the source of standard GR solutions (the seed solutions for the method) and $\Theta_{\mu\nu}$ may contain any new source, for instance new matter fields or a
different gravitational sector. 
Since $\tilde{T}_{\mu\nu}$ must be covariantly conserved, due to the Bianchi identity, $\nabla^{\mu}\Theta_{\mu\nu}=0$ holds. 

By considering the spherically symmetric ansatz
\begin{equation}
ds^2 = e^{\rho(r)}dt^2 -e^{\lambda(r)}dr^2 - r^2 d\sigma^2,
\end{equation}
where $d\sigma^2 = d\theta^2 + \sin^2\theta d\varphi^2$, the Einstein's equations reads
\begin{align}
	\label{eqn:E1}
	\kappa^2 ({T_{0}}^{0}+{\Theta_{0}}^{0}) &= \frac{1}{r^2} - e^{-\lambda}\left(\frac{1}{r^2}-\frac{\lambda'}{r}\right),\\
		\label{eqn:E2}
	\kappa^2 ({T_{1}}^{1}+{\Theta_{1}}^{1})&=\frac{1}{r^2} - e^{-\lambda}\left(\frac{1}{r^2}+\frac{\rho'}{r}\right)\\
		\label{eqn:E3}
	\kappa^2 ({T_{2}}^{2}+{\Theta_{2}}^{2})&= - \frac{e^{-\lambda}}{4}\left(2\rho'' + \rho'^2 - \lambda'\rho' + 2\frac{\rho' - \lambda'}{r}\right).
\end{align}
In the absence of $\Theta_{\mu\nu}$, the solution of the above set is given by 
\begin{equation}
\label{eqn:seed}
ds^2 = e^{\xi(r)}dt^2 - e^{\gamma(r)}dr^2 - r^2 d\sigma^2,
\end{equation}
where 
\begin{equation}
e^{-\gamma(r)}\equiv 1 - \frac{\kappa^2}{r}\int_{0}^{r}x^2 {T_{0}}^{0}(x)dx = 1 - \frac{2 m(r)}{r},
\end{equation}
with $m(r)$ being the usual Misner-Sharp mass function. The gravitational decoupling consists in the assumption that the additional part of the stress-energy tensor $\Theta_{\mu\nu}$ results in a deformation in the seed metric~\eqref{eqn:seed}, given by
\begin{align}
\label{eqn:def1}
&\xi \rightarrow \rho = \xi + \alpha q,
\\
\label{eqn:def2}
&e^{-\gamma} \rightarrow e^{-\lambda}= e^{-\gamma} + \alpha h, 
\end{align}
where $\alpha$ is the deformation parameter and $q$ and $h$ are functions of $r$. From this, one sees that $\alpha \rightarrow 0$ leads to the seed solution. 
Assuming that the addition of $\Theta_{\mu\nu}$ leads to \eqref{eqn:def1} and \eqref{eqn:def2}, the Einstein's equations \eqref{eqn:E1}-\eqref{eqn:E3} are separated in two sets. The first is given by the Einstein's equations for $\alpha = 0$, namely
\begin{align}
&\kappa^2 {T_{0}}^{0} = \frac{1}{r^2} - e^{-\gamma}\left(\frac{1}{r^2} - \frac{\gamma'}{r}\right),\\
&\kappa^2{T_{1}}^{1} = \frac{1}{r^2} - e^{-\gamma}\left(\frac{1}{r^2}+\frac{\xi'}{r}\right),\\
&\kappa^2 {T_{2}}^{2} = -\frac{e^{-\gamma}}{4}\left(2\xi'' +\xi'^2 - \gamma'\xi' + 2\frac{\xi'-\gamma'}{r}\right),
\end{align}
which is solved by \eqref{eqn:seed}. The second set consists of a ``quasi-Einstein'' field equation that contains the source $\Theta_{\mu\nu}$, 
\begin{align}
&\kappa^2{\Theta_{0}}^{0} = -\alpha\frac{h}{r^2} - \alpha\frac{h'}{r},\\
&\kappa^2{\Theta_{1}}^{1} + \alpha Z_1 = -\alpha h\left(\frac{1}{r^2} + \frac{\rho'}{r}\right),\\
&\kappa^2{\Theta_{2}}^{2}+\alpha Z_2 = -\alpha\frac{h}{4}\left(2\rho'' +\rho'^2 +2\frac{\rho'}{r}\right) - \alpha\frac{h'}{4}\left(\rho' +\frac{2}{r}\right),
\end{align}
where 
\begin{align}
Z_1 &= \frac{e^{-\gamma}q'}{r},\\
4Z_2 &= e^{-\gamma}\left(2 q'' + q'^2 + \frac{2q'}{r} + 2\xi'q' - \gamma'q'\right).
\end{align}

With a non-zero deformation parameter ($\alpha\neq0$), the matter sector behaves as an effective anisotropic fluid. In order to avoid exotic matter, one can require that the system satisfies some energy conditions. Once a seed solution is provided, solving the quasi-Einstein system under these energy conditions allows one to generate hairy BHs, as we shall discuss in the following.

\subsection{Hairy black holes}
Hereafter, we set $T_{\mu\nu} = 0$, so that the Schwarzschild metric serves as the unique spherically symmetric seed solution. The anisotropic fluid described by $\Theta_{\mu\nu}$ is then required to satisfy at least one energy condition, SEC, DEC or WEC. These conditions generate inequalities that constrain the metric functions, producing distinct classes of solutions~\cite{Ovalle2021,Avalos2023}. 

For instance, by assuming that $\Theta_{\mu\nu}$ satisfies the SEC, the metric functions must conform to~\cite{Ovalle2021}
\begin{equation}
\label{eqn:secsol}
f_S(r)\equiv e^{\rho(r)}=e^{-\lambda(r)} = 1 - \frac{2M + \alpha \ell}{r} +\alpha e^{-r/M},
\end{equation}
where $\alpha$ encodes the deformation strength and $\ell$ provides a length scale.  Such geometry possess an event horizon located at the largest root of
\begin{equation}
\alpha \ell = r_{h} - 2M + \alpha r_{h}e^{-r_{h}/M}.
\end{equation}
Alternatively, by assuming that $\Theta_{\mu\nu}$ satisfies the DEC, the metric functions must conform to~\cite{Ovalle2021}
\begin{equation}
\label{eqn:decsol}
f_D(r)\equiv e^{\rho(r)}=e^{-\lambda(r)} = 1 - \frac{2M+\alpha \ell}{r} + \frac{Q^2}{r^2} - \frac{\alpha M e^{-r/M}}{r},
\end{equation}
that resembles the SEC solution, but with a sort of ``charge'' term. We remark, however, that $Q\propto \alpha$ does not need to be an electric charge. The event horizon of this geometry is located at the largest root of
\begin{equation}
\alpha \ell = r_{h} - 2M +\frac{Q^2}{r_{h}}- \alpha M e^{-r_{h}/M}.
\end{equation}
Finally, by assuming that $\Theta_{\mu\nu}$ satisfies the WEC, the metric functions must conform to~\cite{Avalos2023}
\begin{equation}
\label{eqn:wecsol}
f_W(r)\equiv e^{\rho(r)}=e^{-\lambda(r)} = 1 - \frac{2 M}{r} + \frac{\jmath M}{r^2}+\frac{\jmath M}{r^2}\log\left(\frac{r}{\beta}\right),
\end{equation}
where $\beta$ and $\jmath$ have dimensions of mass, with $\jmath\propto\alpha$. In this case, the event horizon is located at the largest root of
\begin{equation}
\jmath = \frac{r_{h}(2 M -r_{h})}{M\left(1 + \log\left(\frac{r_{h}}{\beta}\right)\right)}.
\end{equation}

All of these solutions exhibit additional parameters that serve as primary hair. In order to gain some intuition about these additional parameters, one can calculate the ADM mass of these spacetimes~\cite{Hawking1996}.

Specifically for the solutions satisfying the SEC and DEC, one obtains $M_{\text{ADM}} = M + {\alpha \ell}/{2}$. It follows that if $M\to 0^{+}$, the metric functions of the SEC and DEC solutions take the form of the Schwarzschild and Reissner-Nordström solutions, respectively, with $M_{\text{ADM}}=\alpha \ell/2$. Conversely, for the solutions satisfying the WEC, the additional parameters $\beta$ and $\jmath$ do not affect the ADM mass, that is given by $M_{\text{ADM}} = M$. In the following we investigate the role of these additional parameters in the scattering process.

\section{Null geodesics}
\label{sec:nullgeod}
One of the key phenomenological aspects of BH physics concerns how these objects scatter particles. Of particular interest is the scattering of null geodesics, that describes, for instance, the motion of light around these objects. From now on, since the discussed hairy solutions satisfy $e^{\rho(r)}=e^{-\lambda(r)}$, it will be convenient to write the line element of the hairy BH spacetimes as
\begin{equation}
\label{eqn:gauge}
ds^2 = f(r)dt^2 - f(r)^{-1}dr^2 - r^2 d\sigma^2,
\end{equation} 
where $f(r) \equiv e^{\rho(r)} = e^{-\lambda(r)}$, whose explicit forms for the SEC-, DEC-, and WEC-satisfying hairy solutions are given by Eqs.~\eqref{eqn:secsol}, \eqref{eqn:decsol}, and \eqref{eqn:wecsol}, respectively.

Light rays follow paths that obey $g_{\mu\nu}\dot{x}^{\mu}\dot{x}^{\nu} = 0$, where the overdot denotes differentiation with respect to an affine parameter. Due to spherical symmetry, let us restrict the motion, without loss of generality, to the equatorial plane ($\theta = \pi/2$). The Lagrangian of a free particle is
\begin{align}
\mathcal{L} &= \frac{1}{2}g_{\mu\nu}\dot{x}^{\mu}\dot{x}^{\nu},\nonumber\\
&=\frac{1}{2}\left(f(r)\dot{t}^2 - f(r)^{-1}\dot{r}^2 - r^2\dot{\varphi}^2\right),
\end{align}
from which one can extract the following conserved quantities
\begin{align}
\label{eqn:energy}
E&=\frac{\partial\mathcal{L}}{\partial\dot{t}} = f(r)\dot{t},\\
\label{eqn:angular}
-L&=\frac{\partial\mathcal{L}}{\partial\dot{\varphi}}=-r^2\dot{\varphi}.
\end{align}
Hence, on the equatorial plane, light rays satisfy
\begin{equation}
\label{eqn:rgeod}
\dot{r}^2 + V_{\text{eff}} = E^2,
\end{equation}
where $V_{\text{eff}}$ is the effective potential, given by
\begin{equation}
\label{eqn:effpotential}
V_{\text{eff}} = E^2b^2 \frac{f(r)}{r^2},
\end{equation}
where $b=L/E$ is the so-called impact parameter. 

Equation~\eqref{eqn:rgeod} allows for closed circular orbits, known as light rings. These orbits satisfy the conditions $\dot{r} = 0$ and $\ddot{r}=0$, which means that the location of the light rings, say $r=r_c$, are critical points of the effective potential, namely $V'_{\text{eff}}(r_c)= 0$. If $r_c$ represents a local maximum of the effective potential, the light ring is said to be unstable. Otherwise, if $r_c$ represents a local minimun of the effective potential, the light ring is said to be stable. The conditions on $\dot{r}$ and $\ddot{r}$ yield
\begin{align}
b_{c} = \frac{r_c}{\sqrt{f(r_c)}},\quad r_c =\frac{2 f(r_c)}{f'(r_c)},
\end{align}
where $b_{c}$ is the critical impact parameter associated to the light ring.

\begin{figure}[!h]
\centering
\includegraphics[width=\linewidth]{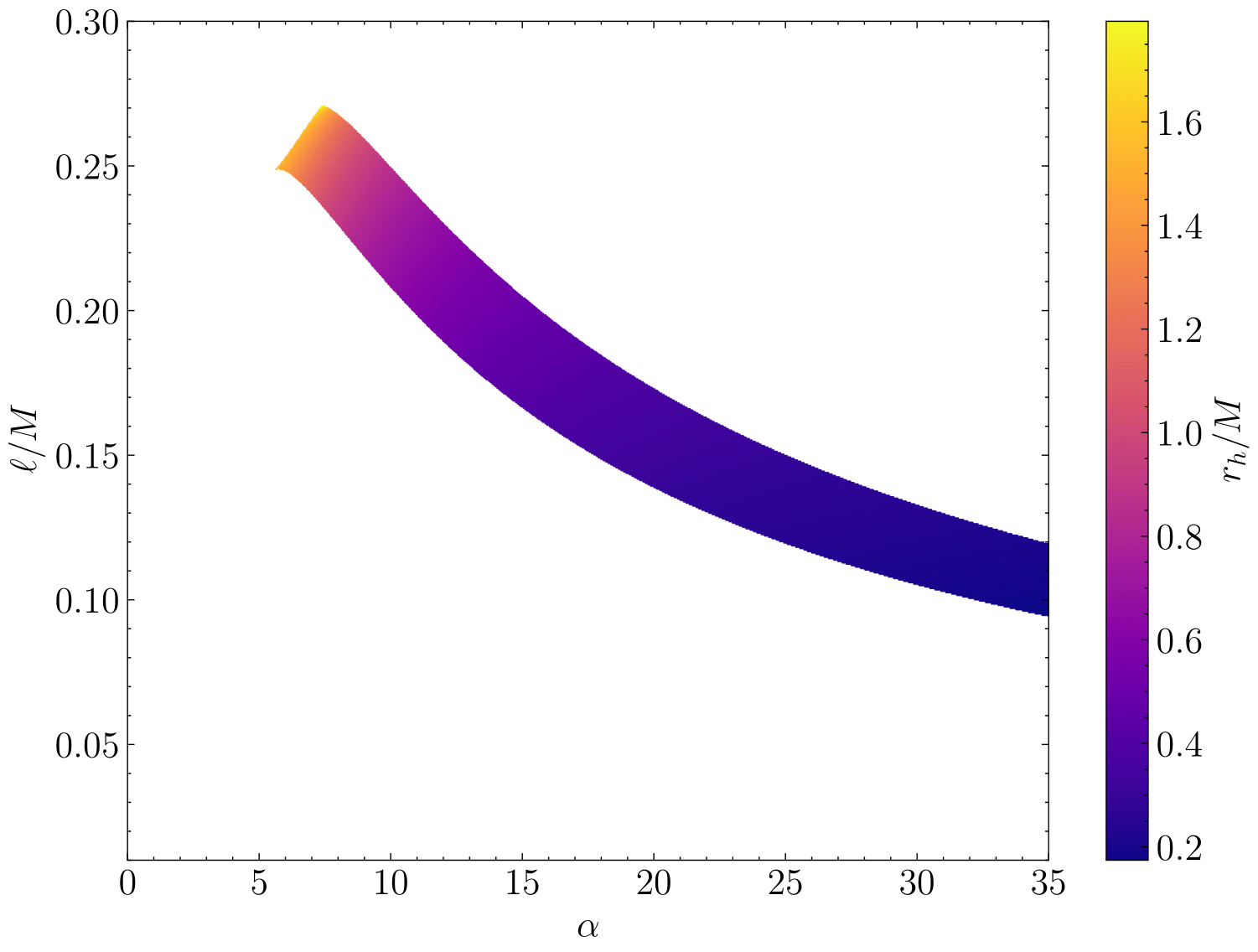}
\includegraphics[width=\linewidth]{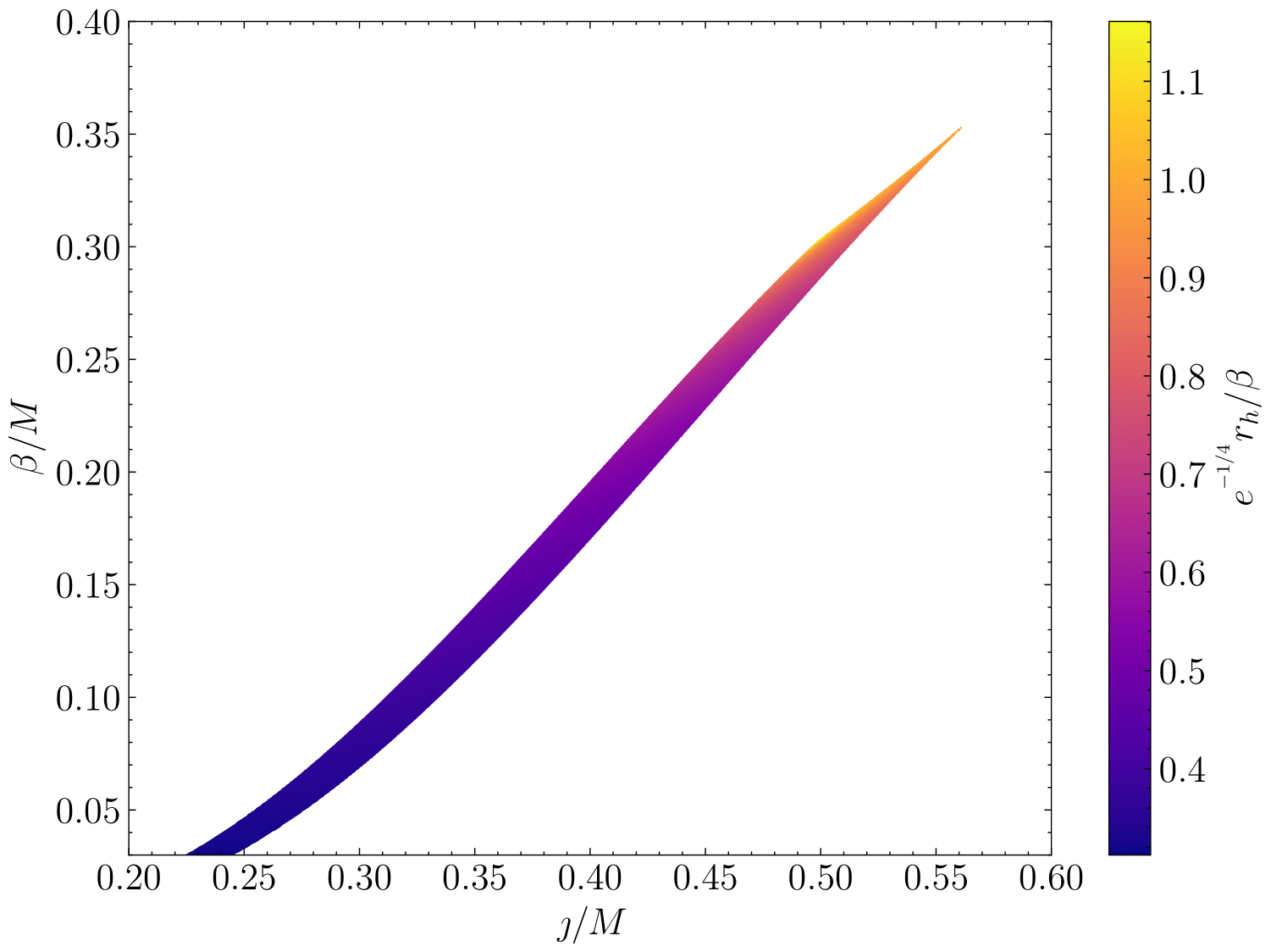}
\caption{Regions of the parameter space where the effective potential develops a local minimum outside the event horizon. The top panel corresponds to the SEC solution, while the bottom panel shows the WEC solution. The color coding indicates whether the corresponding energy condition is satisfied throughout the region exterior to the event horizon, helping to visualize possible violations in specific domains.}
	\label{fig:params-well}
\end{figure}

We point out that stable light ring structures are often found in exotic horizonless compact objects \cite{Macedo2013, Delhom2019, Junior2020}. However, certain parameter regimes of the considered hairy BH solutions may also permit them. We remark that other examples of BHs with multiple light rings are known (see Ref.~\cite{guo2023black} for a comprehensive analysis of their physical viability). Such structures have been shown to give rise to distinctive optical appearances (see, for instance, Refs.~\cite{gan2021photon1, gan2021photon2}). Since the transcendental nature of the metric functions prevents full analytical exploration, we employ a numerical scheme that systematically scans the region of the parameter space that could allow for multiple light rings, including stable ones. 
In Fig.~\ref{fig:params-well}, we display the regions of the parameter space, corresponding to the SEC and WEC solutions, within which the effective potential exhibits a local minimum outside the event horizon. These regions are color-coded according to the event horizon location $r_h$, suitably normalized. In our numerical runs we did not find any DEC solution with stable light rings in the parameter ranges analyzed.

\begin{figure}[!h] 
	\centering
	\includegraphics[width=\columnwidth]{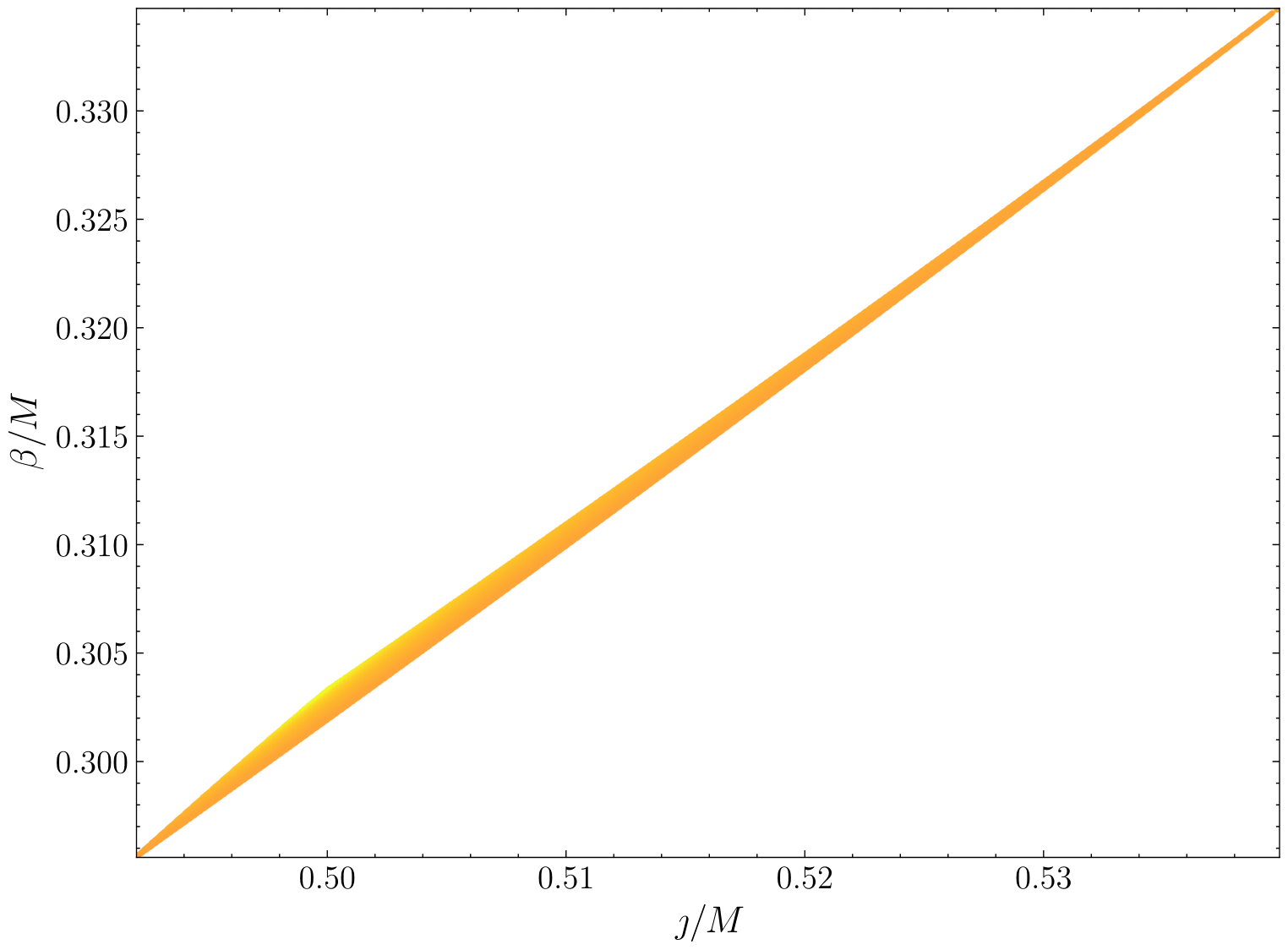}
	\caption{Region of the parameter space of the WEC (satisfied everywhere outside the event horizon) solution in which the effective potential presents a well.}
	\label{fig:WEC-satisfied}
\end{figure}

For the solution given by Eq.~\eqref{eqn:secsol}, it is known that the SEC holds only for $r \geq 2M$~\cite{Ovalle2021}. The top panel of Fig.~\ref{fig:params-well} shows that  all configurations exhibiting a local minimum in the effective potential have their event horizons located at $r_h < 2M$, and therefore, in configurations with stable light rings, the SEC is unavoidably violated outside the event horizon in the region $r_h<r<2M$. In contrast, for the solution~\eqref{eqn:wecsol}, the WEC is satisfied for $r\geq \beta e^{1/4}$~\cite{Avalos2023}, and the bottom panel of Fig.~\ref{fig:params-well} shows that one is able to find configurations with stable light rings that satisfy the WEC on and outside the event horizon. This region is outlined in Fig.~\ref{fig:WEC-satisfied}. It is worth noting that, although there are configurations endowed with stable light rings that satisfy the WEC on and outside the event horizon (see Fig.~\ref{fig:WEC-satisfied}), all configurations present in the bottom panel of Fig.~\ref{fig:params-well} violate the SEC in some region outside the horizon. 

\begin{figure}[ht!]
	\centering
	\includegraphics[width=\columnwidth]{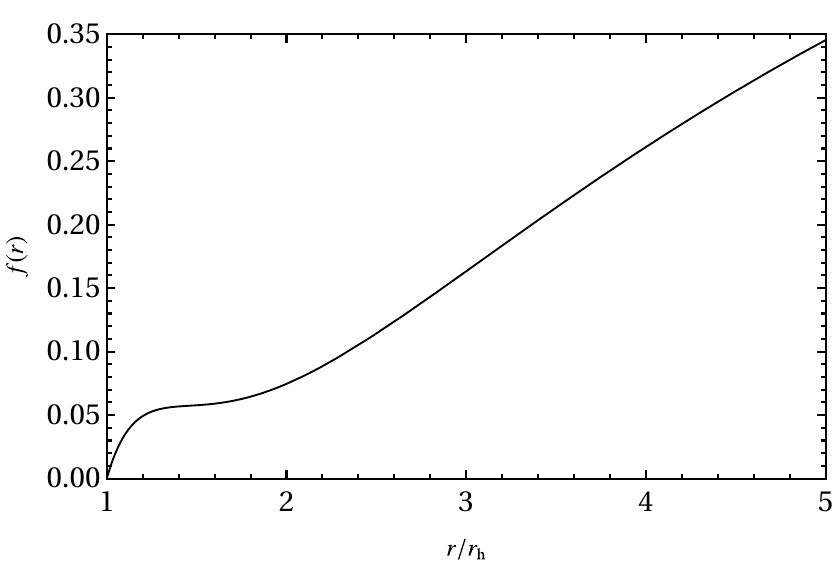}
	\vspace{0.2cm}
	\includegraphics[width=\columnwidth]{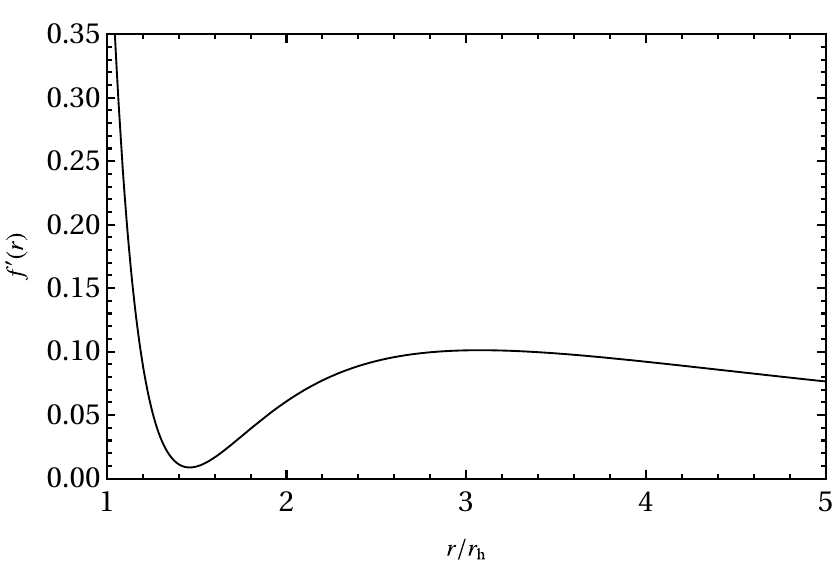}
	\vspace{0.2cm}
	\includegraphics[width=\columnwidth]{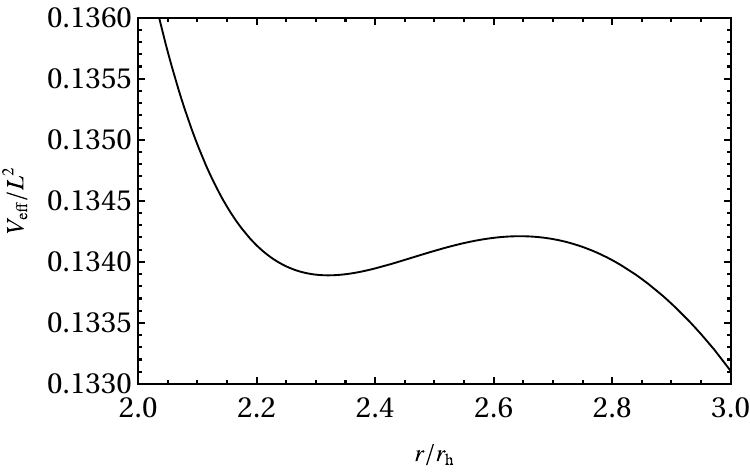}
	\caption{WEC hairy BH configuration with monotonically increasing redshift function outside the horizon and stable light ring structure ($\jmath/M = 0.534,~\beta/M = 0.325$). The top panel shows the redshift function, while the middle panel shows its derivative, making explicit the monotonicity of $f(r)$. The bottom panel shows the occurrence of a local minimum, even though $f(r)$ is strictly increasing.}
	\label{fig:monotonic}
\end{figure}

In Ref.~\cite{Cvetivc2016} it is conjectured that, in spherically symmetric spacetimes, the violation of either the DEC or SEC is a necessary condition for the existence of stable light rings. Our results are consistent with this conjecture, since every configuration we found to be endowed with a stable light ring is associated with the violation of the SEC. It is worth noting that the physical viability of BHs with multiple light rings was thoroughly examined in Ref.~\cite{guo2023black}, where a counterexample to the aforementioned conjecture was identified. That study, nonetheless, indicates that the conjecture still holds when the metric satisfies $g_{tt}g_{rr}=-1$, which is the case we are considering here.  Furthermore, it was shown in Ref.~\cite{Cvetivc2016} that the SEC implies that the redshift function, $|g_{tt}|$,  must be monotonically increasing. One might therefore assume that the existence of stable light rings is necessarily associated to non-monotonic behavior of the metric function outside the event horizon. However, this assumption does not hold in general. Within the class of configurations analyzed in this work, we identify explicit counterexamples to this hypothesis.  In Fig.~\ref{fig:monotonic} we plot one such counterexample, in which although $f(r)$ (cf. Eq.~\eqref{eqn:gauge}) is monotonically increasing outside the horizon and the effective potential exhibits a local minimum, hence a stable light ring is present in the spacetime.


\section{Absorption Cross section}
\label{sec:absorption}
\subsection{Partial-waves approach}

The dynamics of a massless scalar field $\Phi$ in the background of the hairy BH solutions is given by the Klein-Gordon equation, 
\begin{equation}
\frac{1}{\sqrt{-g}}\partial_{\mu}(\sqrt{-g}g^{\mu\nu}\partial_{\nu}\Phi) = 0,
\end{equation}
where $g$ is the determinant of the metric tensor $g_{\mu\nu}$. Due to spherical symmetry, the Klein-Gordon equation can be appropriately separated using the ansatz   
\begin{equation}
\Phi_{lm\omega}(t, r,\theta,\varphi) = \frac{\psi_{l\omega}(r)}{r} e^{-i\omega t} Y_{l m}(\theta,\varphi),
\end{equation} 
with $Y_{l \omega}$ being the spherical harmonics. 
Since $r^2\nabla^2 Y_{l \omega} = -l(l+1)Y_{l \omega}(\theta,\varphi)$, the radial equation of the scalar field is given by
\begin{equation}
\label{eqn:KG1}
	f(r)\frac{d}{dr}\left(f(r)\psi_{l \omega}'(r)\right) + \left(\omega^2 - V_{l}\right)\psi_{l \omega}(r) = 0,
\end{equation}
where $V_{l}(r)$ is the scattering potential, 
\begin{equation}
\label{eqn:V-scatt}
V_{l}(r)= f(r)\left(\frac{l(l+1)}{r^2}+\frac{f'(r)}{r}\right).
\end{equation}
Introducing the tortoise coordinate, implicitly defined by $x = \int f(r)^{-1}dr$, the radial equation~\eqref{eqn:KG1} can be rewritten in a Schrödinger-like form, namely
\begin{center}
\begin{equation}
\label{eqn:schr}
\left[\frac{d^2}{dx^2} + \omega^2 - V_{l}(x)\right]\psi_{l \omega}(x) = 0.
\end{equation}
\end{center}
In this new coordinate system, the range $r\in (r_{h},\infty)$ is mapped onto $x \in (-\infty, \infty)$, where $r_{h}$ is the geometrical locus of the event horizon.

\begin{figure}[!h] 
	\centering
	\includegraphics[width=\columnwidth]{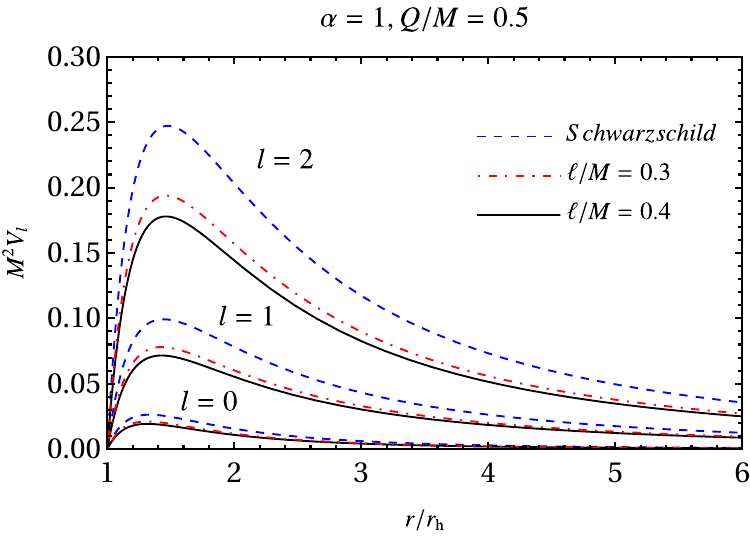}
	\vspace{0.5em} 
	\includegraphics[width=\columnwidth]{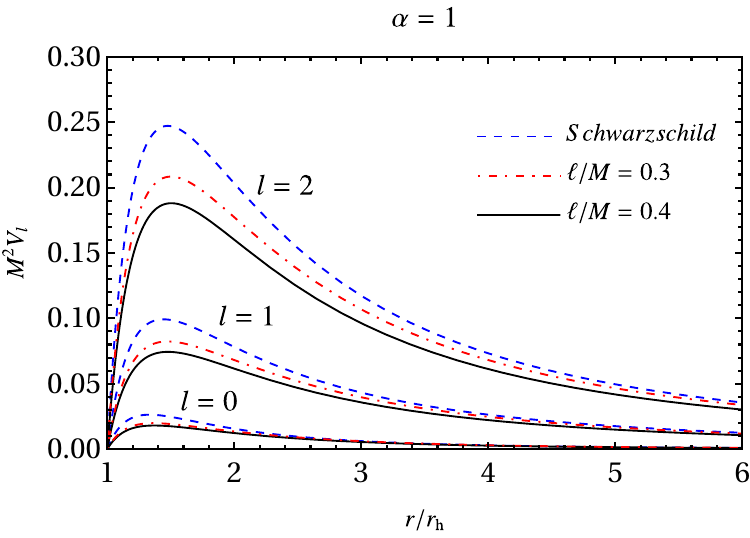}
	\vspace{0.5em}
	\includegraphics[width=\columnwidth]{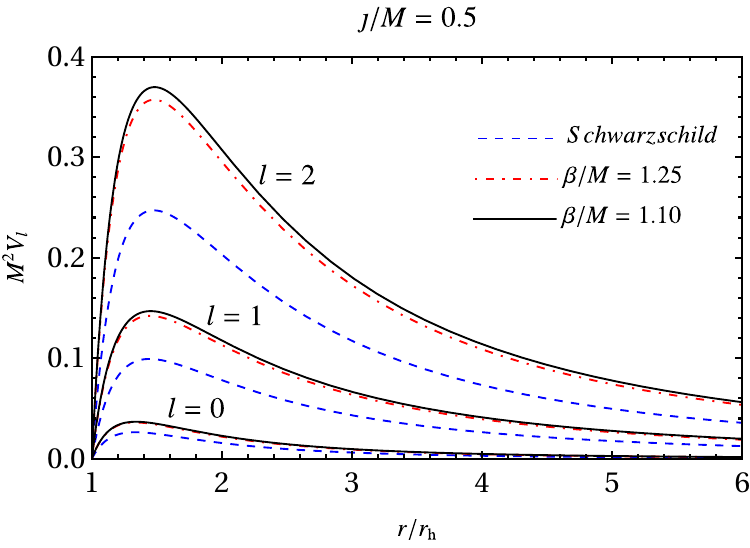}
	\caption{Scattering potential as functions of the radial coordinate (in units of the event horizon radius, $r_\text{h}$), all with $\alpha = 1$ (or $\alpha/M = 0.5$ in the WEC case). The top, middle, and bottom panels correspond to the DEC, SEC and WEC BH configurations, respectively. In all cases, the Schwarzschild scattering potential is included for comparison.}
	\label{fig:scalar-potentials}
\end{figure}

In Fig.~\ref{fig:scalar-potentials} we plot the scattering potential of the hairy BHs obtained via the gravitational decoupling for some values of the solution's parameters. Specifically, in the top panel of Fig. \ref{fig:scalar-potentials} we plot the scattering potential of the DEC solution, which is smaller than the Schwarzschild potential and decreases as $\ell$ increases. The same behavior occurs in the SEC solution, depicted in the middle panel of Fig.~\ref{fig:scalar-potentials}. For the WEC solution ($\beta>0$), the scattering potential is larger than in the corresponding Schwarzschild case, as illustrated in the bottom panel of Fig.~\ref{fig:scalar-potentials}.

We analyze scalar waves obeying scattering boundary conditions, where the radial wave function must satisfy specific asymptotic behavior, namely
\begin{equation}
\label{eqn:boundary}
\Phi_{l \omega} \sim \begin{cases}
	T_{\omega l} e^{-i\omega r},& r\rightarrow r_{h}\,\,(x \rightarrow -\infty), \\
	e^{-i\omega r} + R_{\omega l} e^{i\omega r},& r\rightarrow\infty\,\, (x\rightarrow\infty), 
\end{cases}
\end{equation}
where $|T_{\omega l}|^2$ and $|R_{\omega l}|^2$ denote the transmission and reflection coefficients, respectively. The quantity $|T_{\omega l}|^2$ represents the probability of absorption of the wave, while $|R_{\omega l}|^2$ corresponds to the probability of scattering~\cite{Futterman1988}. The flux conservation law leads to
\begin{equation}
|T_{\omega l}|^2 + |R_{\omega l}|^2 = 1.
\end{equation}
One can then compute the total scalar absorption cross section as a sum over the partial absorption cross sections, namely
\begin{equation}
\label{eqn:totcross}
\sigma(\omega) = \sum_{l=0}^{\infty}\sigma_l,
\end{equation}
where the partial absorption cross sections are given by 
\begin{equation}
\sigma_l=\frac{\pi}{\omega^2}(2l+1)|T_{\omega l}|^2.
\end{equation}

Notably, in the absence of potential wells, the total scalar absorption cross section of stationary BHs possesses two well-known limits. In the low-frequency regime, it is known that the absorption cross section of massless scalar fields tends to the surface area of the event horizon~\cite{Das1997,Higuchi2001, Higuchi2002}, namely
\begin{align}
	\lim_{\omega\to 0}\sigma(\omega)= A_{\text{h}} &= \int_{0}^{\pi}\int_{-\pi}^{\pi}\sqrt{g_{\theta\theta}g_{\varphi\varphi}}d\varphi d\theta\big|_{r=r_h}= 4\pi {r_h}^2.
\end{align}
Furthermore, in the high-frequency regime, the absorption cross section exhibits an oscillatory behavior, which was firstly obtained by Sanchez, through a numerical fit of the absorption cross section of the Schwarzschild BH~\cite{Sanchez1978}. Later, it was shown that this oscillatory behavior is a universal feature of the absorption~\cite{Decanini2011}, and is described by the sinc approximation, namely
\begin{equation}
\sigma_{\text{hf}} = \left[1-\frac{8\pi\lambda_c}{\Omega_c}e^{-\frac{\pi\lambda_c}{\Omega_c}}\text{sinc}\left(\frac{2\pi\omega}{\Omega_c}\right)\right]\sigma_{\text{geo}},
\end{equation}
where $\sigma_{\text{geo}} = \pi b_{c}^2$ is the geometric capture cross section, 
$\Omega_c$ is the angular velocity of null geodesics at the unstable light-ring, given by
\begin{equation}
\Omega_c = \frac{\sqrt{f(r_c)}}{r_c},
\end{equation}
and $\lambda_c$ is the Lyapunov exponent, which measures the average rate at which neighboring trajectories converge or diverge in phase space \cite{Cardoso2009}, and it is given by
\begin{equation}
\lambda_c = \frac{1}{\sqrt{2}{r_{c}}}\sqrt{f(r_c)\left(2 f(r_c) - {r_{c}^2} f''(r_c)\right)}.
\end{equation}
\subsection{Quasibound states}
Remarkably, one can find instances in which the scattering potential of these hairy BH solutions presents a local minimum. The presence of these potential wells gives rise to quasibound states, which are associated to long-lived modes characterized by complex frequencies with a small imaginary part. 

In Sec.~\ref{sec:nullgeod}, the existence of stable null geodesics was analyzed, and a conjecture relating their occurrence and the violation of the SEC was mentioned. 
Here, let us derive a sufficient condition for the existence of a local minimum in the scattering potential, at least for $l=0$, and, hence, for the occurrence of quasibound states. Specifically, by considering a spacetime given by Eq.~\eqref{eqn:gauge}, if $f(r)$ is not monotonically increasing and $0 \leq f(r) < 1$ in the region outside the horizon, the scattering potential presents a local minimum at least for $l=0$.


To understand why this is true, let us consider the expression of the scattering potential for $l=0$, namely
\begin{equation}
V_{0}(r) = \frac{f(r)f'(r)}{r}.
\end{equation}
If $f(r)$ is not a strictly increasing function outside the horizon, there exists a region in which $f'(r)<0$, and, therefore, within this region $V_{0}(r)<0$. But, since $f(r_{h})=0$ and $\lim_{r\to\infty}f'(r)=0^{+}$, it follows that $\lim_{r\to\infty}V_{0}(r)=0^{+}$. Therefore, $V_{0}$ exhibits at least one local minimum within the region where it is negative. 

One remark about this result is that, given a spacetime with line element~\eqref{eqn:gauge}, the SEC implies
\begin{equation}
f''(r) + \frac{2 f'(r)}{r}\geq0,
\end{equation}
and if $f'(r)<0$ within a region, with $\lim_{r\to\infty}f'(r)=0^{+}$, it follows that the SEC must be violated in a finite region outside the horizon. We note that if $\ell\neq 0$ the analytical treatment is not easily accessible, much like in the case of the effective potential. In such cases, a similar numerical investigation is necessary.

\begin{figure}[ht!]
	\centering
	\includegraphics[width=\columnwidth]{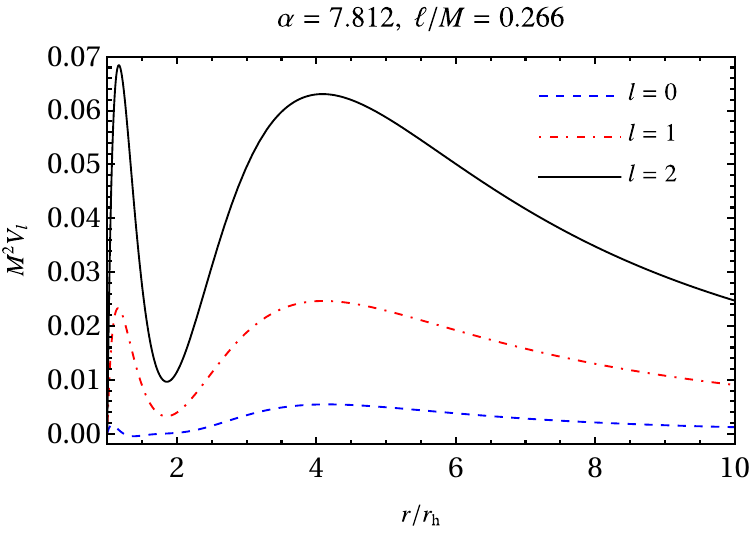}
	\vspace{0.2cm}
	\includegraphics[width=\columnwidth]{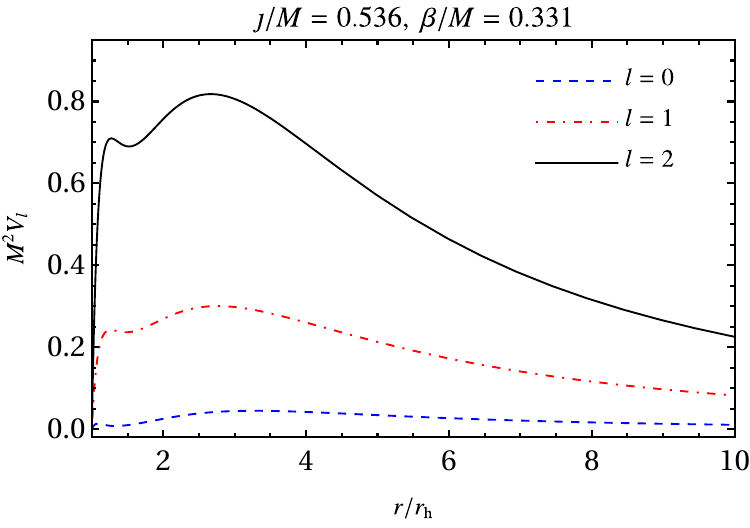}
	\vspace{0.2cm}
	\includegraphics[width=\columnwidth]{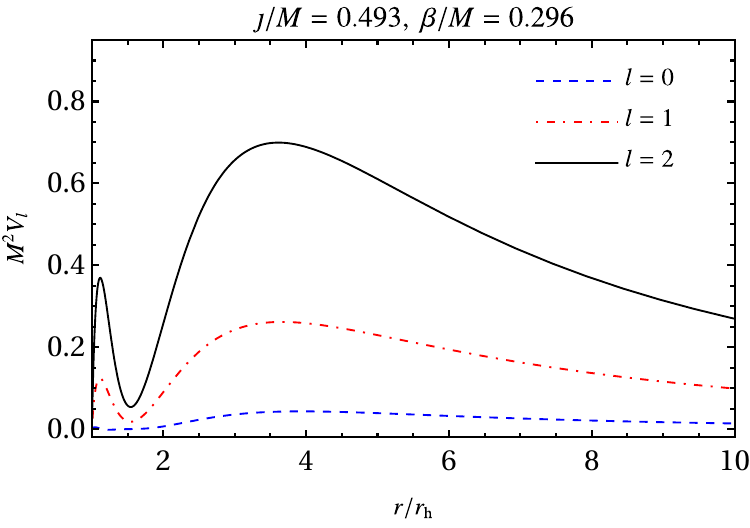}
	\caption{Potential well for chosen configurations of the SEC (top panel) and WEC (middle and bottom panel) solutions for $l=\{0,1,2\}$, as a function of $r$, in units of $r_{h}$.}
	\label{fig:potential-eco}
\end{figure}

\begin{figure}[!h] 
	\centering
	\includegraphics[width=\columnwidth]{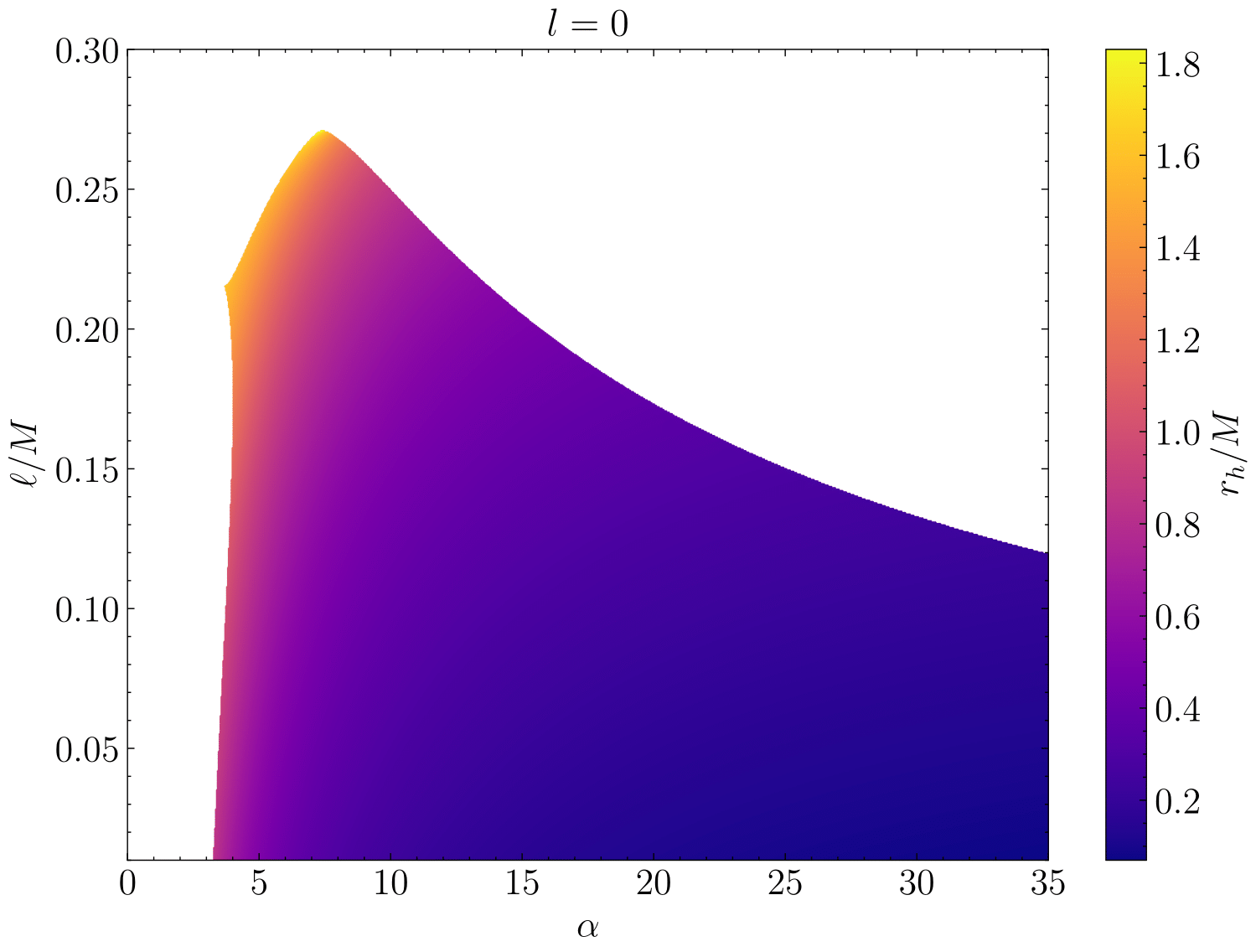}
	\vspace{0.5em} 
	\includegraphics[width=\columnwidth]{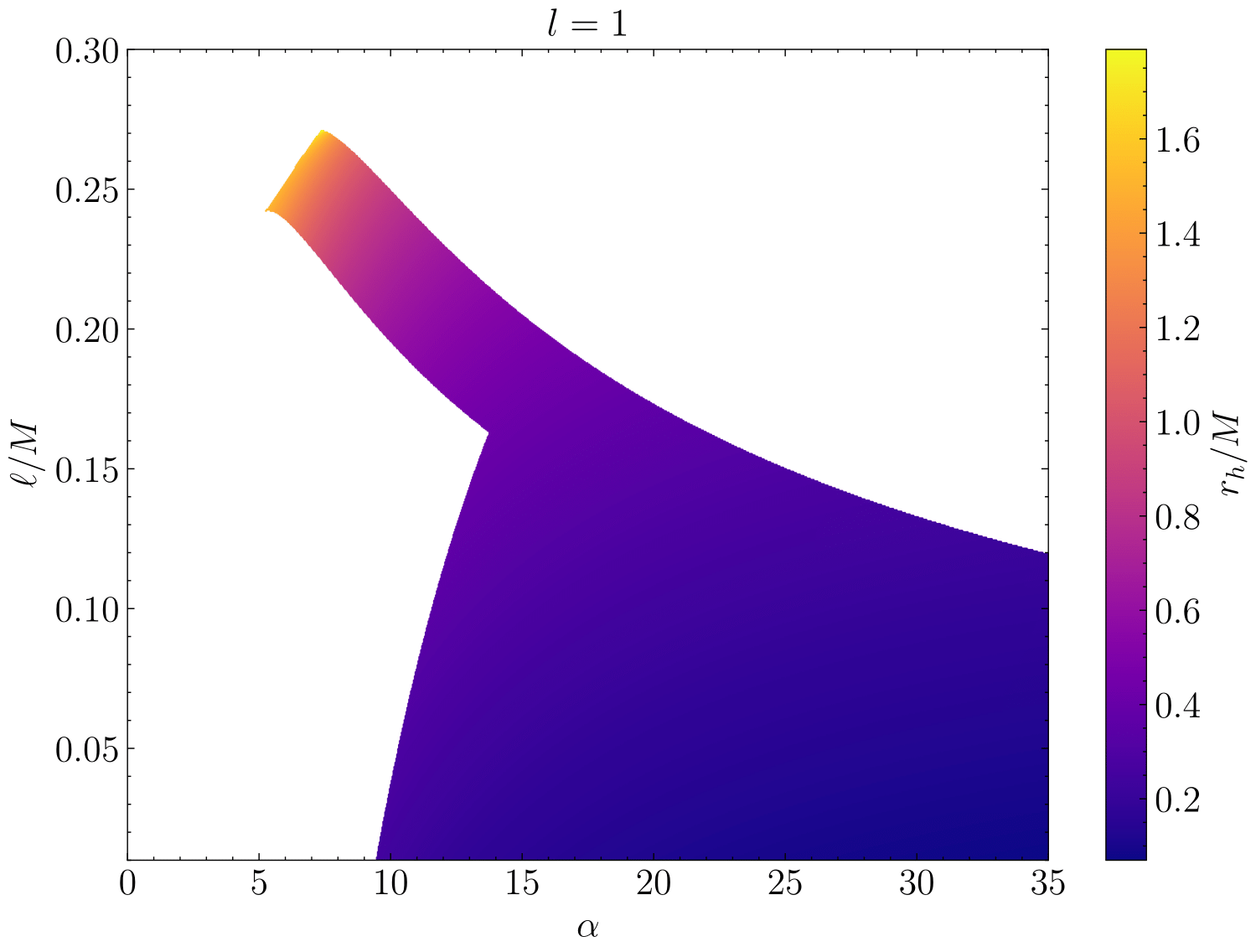}
	\vspace{0.5em}
	\includegraphics[width=\columnwidth]{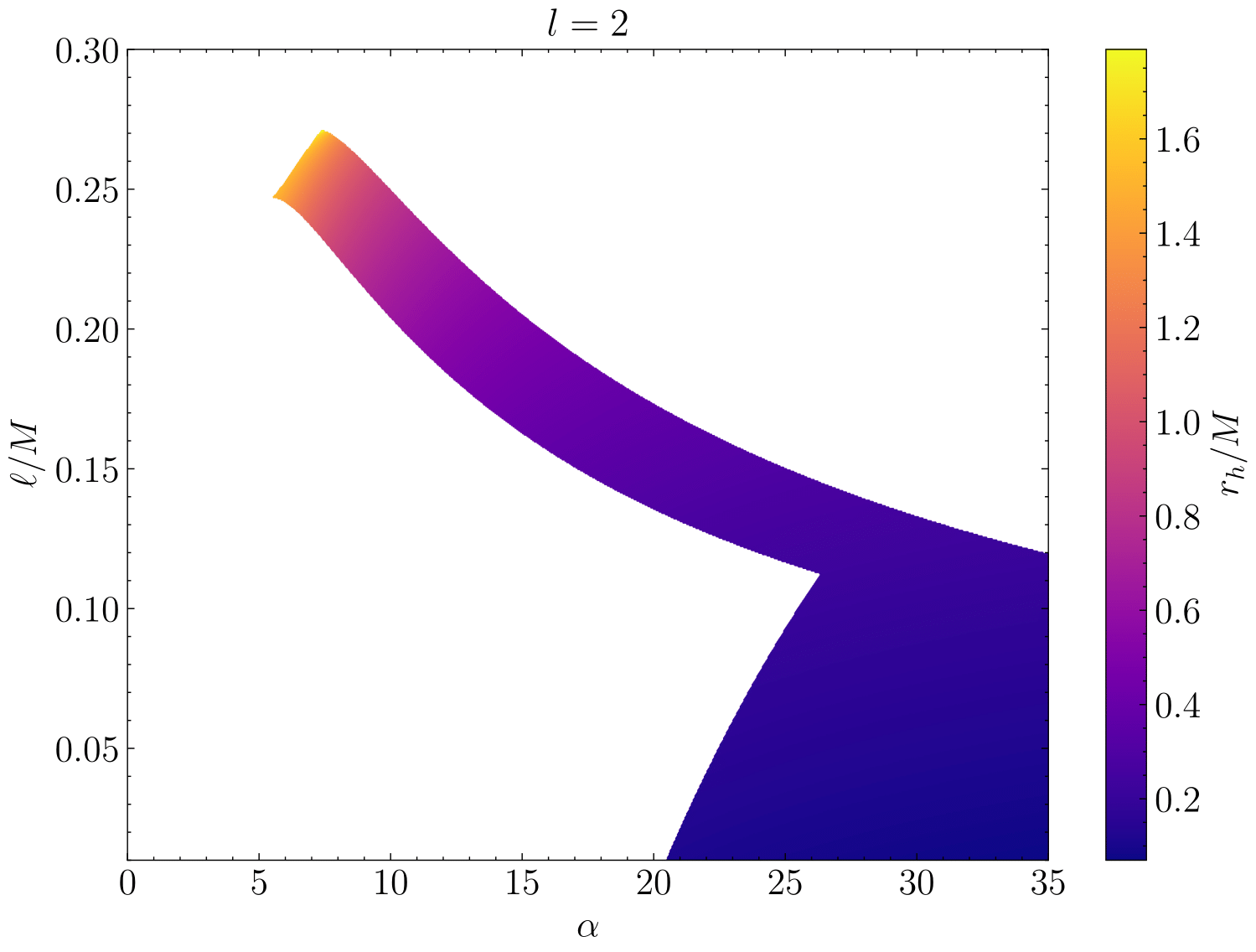}
	\caption{Regions of the parameter space of the SEC solution where the scattering potential presents a local minimum outside the event horizon, for different values of $l$.}
	\label{fig:SEC-scat-l}
\end{figure}

\begin{figure}[!h] 
	\centering
	\includegraphics[width=\columnwidth]{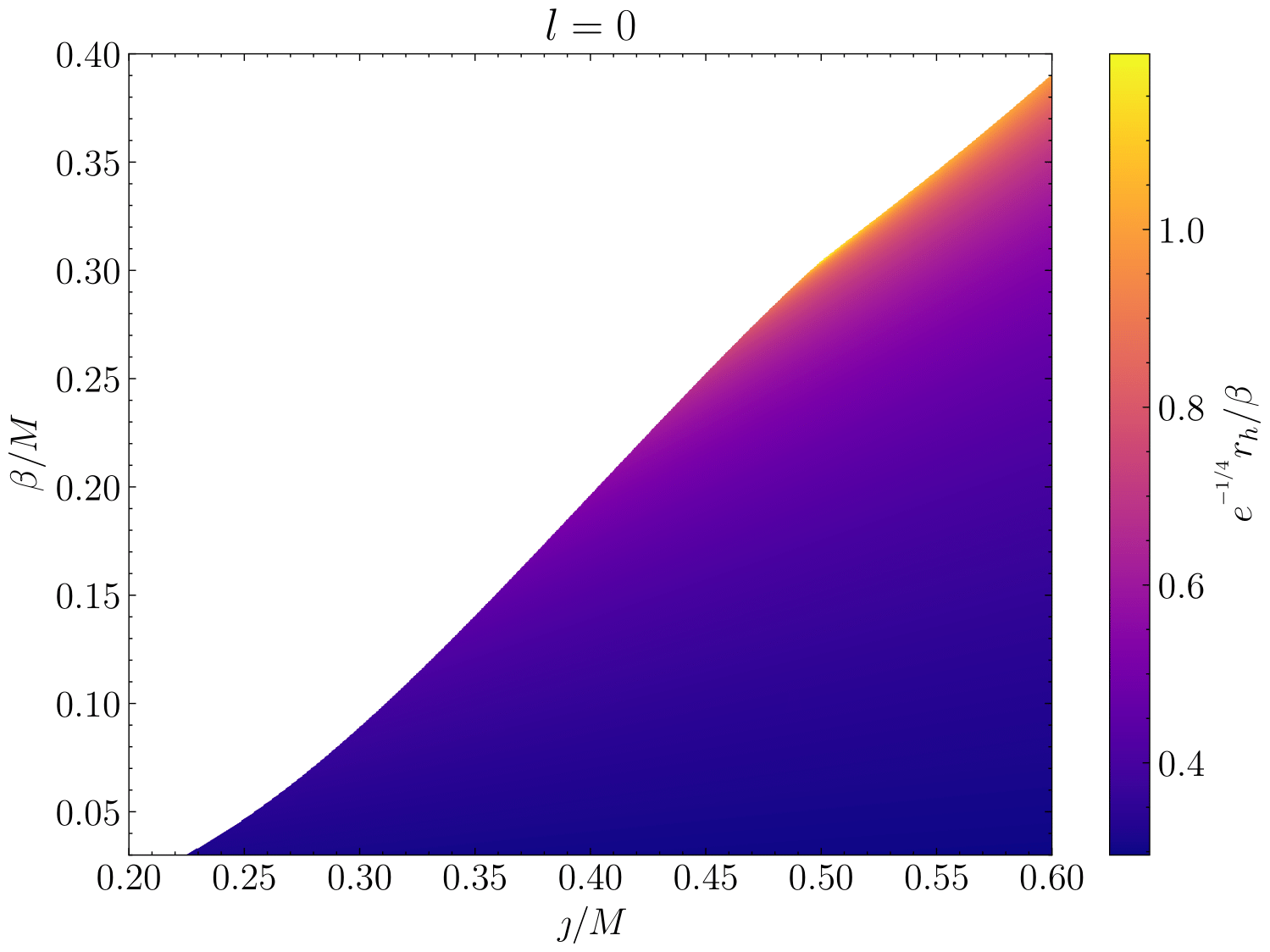}
	\vspace{0.5em} 
	\includegraphics[width=\columnwidth]{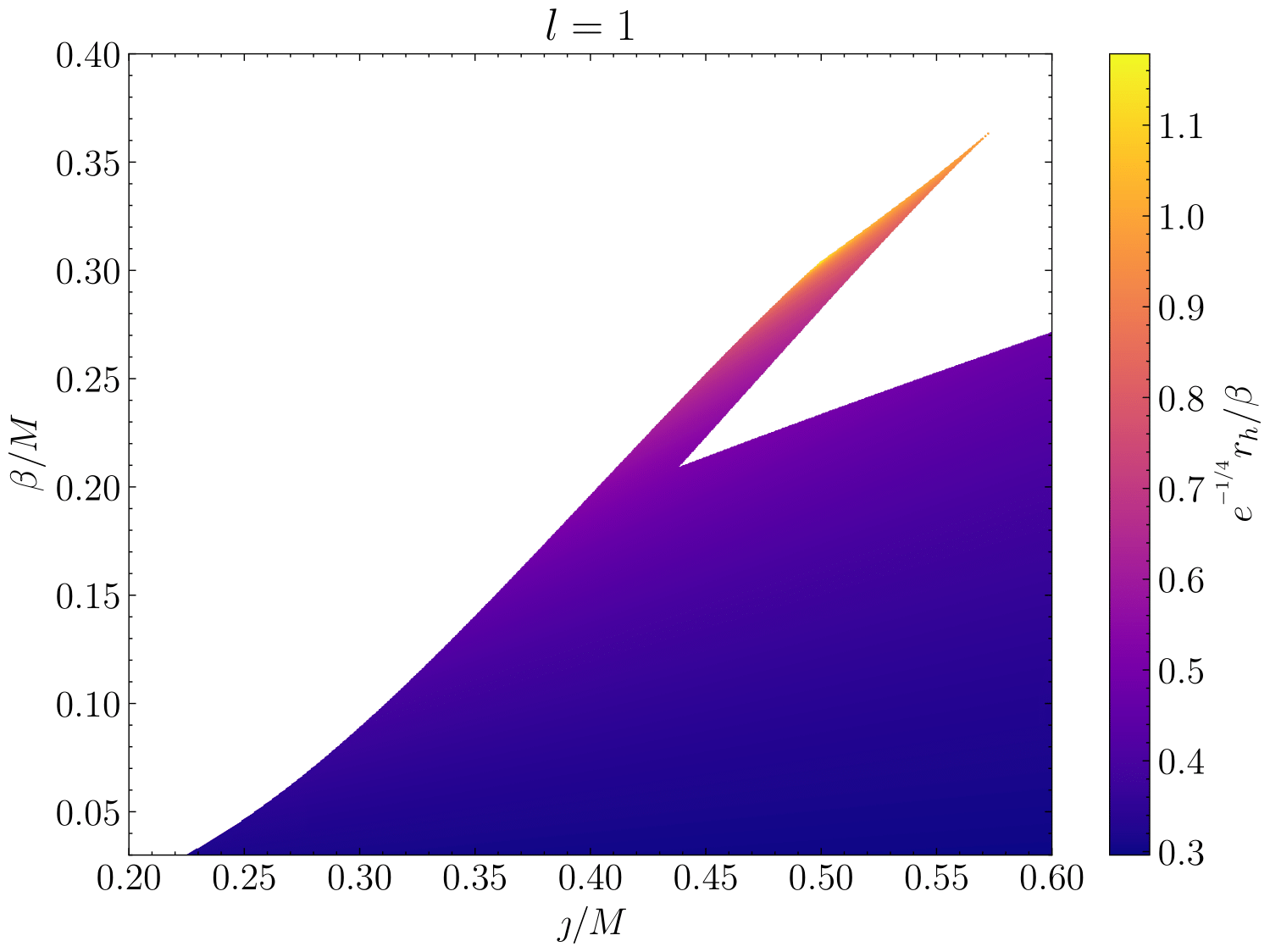}
	\vspace{0.5em}
	\includegraphics[width=\columnwidth]{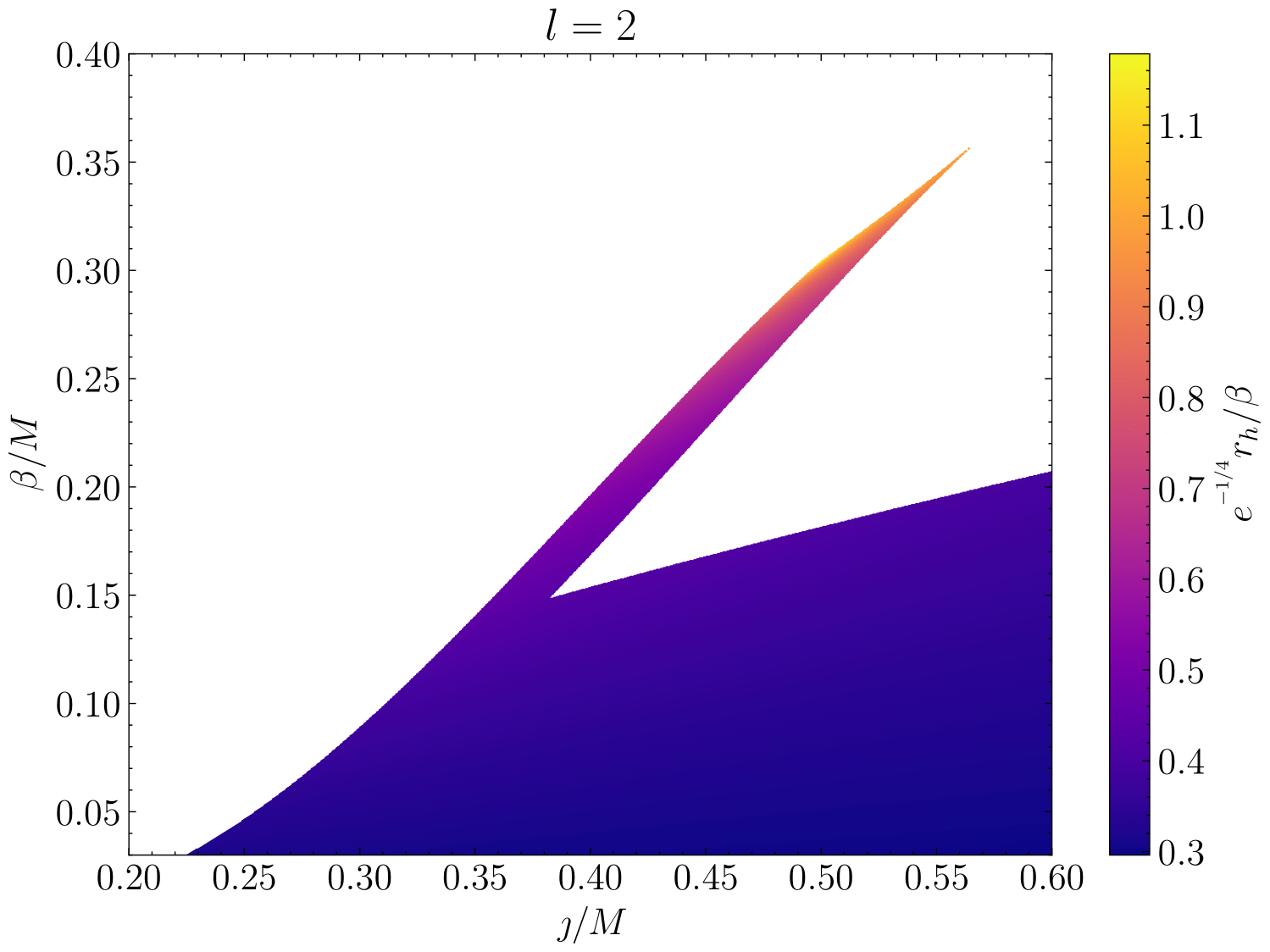}
	\caption{Regions of the parameter space of the WEC solution where the scattering potential presents a local minimum outside the event horizon, for different values of $l$.}
	\label{fig:WEC-scat-l}
\end{figure}

In Fig.~\ref{fig:potential-eco} we plot some scattering potentials of SEC and WEC hairy BH solutions that exhibit a local minimum. The WEC configurations were chosen from the set shown in Fig.~\ref{fig:WEC-satisfied}.
In Figs.~\ref{fig:SEC-scat-l} and~\ref{fig:WEC-scat-l} we show the regions of the parameter space in which the scattering potential presents a local minimum outside the event horizon for different values of $l$ for the SEC and WEC hairy BH solutions, respectively. One can see from Figs.~\ref{fig:SEC-scat-l} and~\ref{fig:WEC-scat-l} that there is no configuration in which a potential minimum occurs such that the SEC is satisfied on and outside the event horizon. 

The presence of quasibound states may lead to distinct signatures in the transmission coefficients, resulting in absorption profiles remarkably different from the standard BHs of GR. Specifically, the absorption spectra may present narrow spectral lines (resonant absorption peaks) located at the real part of the quasibound frequencies. Such a signal is commonly found in wormhole spacetimes~\cite{Junior2020,magalhaes2023asymmetric,furuta2024spectral}, in weakly dissipative ultracompact stars~\cite{Macedo2018}, as well as in BH remnants within modified gravity theories~\cite{Delhom2019}. Furthermore, trapped modes are associated with the arising of echoes in the ringdown profile~\cite{Cardoso2017,guo2022quasinormal,guo2022echoes}. Such behavior has already been reported as existing in the time profile of scalar perturbations in some hairy BH solutions obtained via the gravitational decoupling~\cite{Yang2024} (for a comprehensive analysis of this topic, see Ref.~\cite{yang2025qnm}).

To compute the frequencies associated with the quasibound states, one may employ the method of direct integration, proposed in Ref.~\cite{Chandrasekhar1975}. Such modes obey the boundary conditions given by
\begin{equation}
\label{eqn:trappedboundary}
\Phi_{l \omega} \sim \begin{cases}
	e^{i\omega x},& x \rightarrow \infty, \\
	e^{-i\omega x} ,& x\rightarrow -\infty. 
\end{cases}
\end{equation}
Under these conditions, one solves Eq.~\eqref{eqn:schr} in the two asymptotic regions up to a chosen matching point. The requirement that the two solutions coincide at this point is satisfied only for discrete frequency values, which correspond to the trapped modes. For analyses of quasinormal modes of hairy BH solutions generated via gravitational decoupling with single-peaked scattering potentials, see Refs.~\cite{cavalcanti2022echoes, yang2023probing}.

We remark that, in the vicinity of the resonant peaks in the transmission coefficient, an approximation based on the Breit-Wigner expression for nuclear scattering is valid~\cite{Macedo2018}, namely
\begin{equation}
	\label{eqn:breit-wigner}
	{|T_{\omega l}|^{2}}\Big|_{\omega\approx\omega_{r}} = \frac{B_{\omega l}}{(\omega-\omega_{R})^2 + \omega_{I}^2},
\end{equation}
where $B_{\omega l}$ is a constant and $\omega_{R}$ and $\omega_{I}$ are respectively the real and imaginary parts of the modes. The function~\eqref{eqn:breit-wigner} has a local maximum at $\omega = \omega_{R}$, which determines the position of the resonant peaks, while $\omega_{I}$, which is related to the damping rate of the mode~\cite{kokkotas1999quasi}, determines the sharpness and width of the peaks. Thus, by computing the transmission coefficient through direct integration, one can perform a nonlinear fit of Eq.~\eqref{eqn:breit-wigner} to the numerical data and extract the imaginary part $\omega_{I}$ \cite{Berti2009} (see also Refs.~\cite{Macedo2018, Delhom2019}). 

\section{Results}
\label{sec:results}
In this section we present a selection of numerical results for the absorption of massless scalar fields by hairy BH solutions obtained via gravitational decoupling. To achieve this, we numerically integrate Eq.~\eqref{eqn:schr}, imposing the boundary conditions specified in Eq.~\eqref{eqn:boundary}. This process involves constructing a numerical tortoise coordinate and carefully implementing boundary conditions at both the numerical horizon (positioned very near the event horizon) and numerical infinity (placed at a sufficiently large distance). Fine-tuning the location of numerical infinity is necessary to achieve the desired precision.

\begin{figure}[ht!]
	\centering
	\includegraphics[width=\columnwidth]{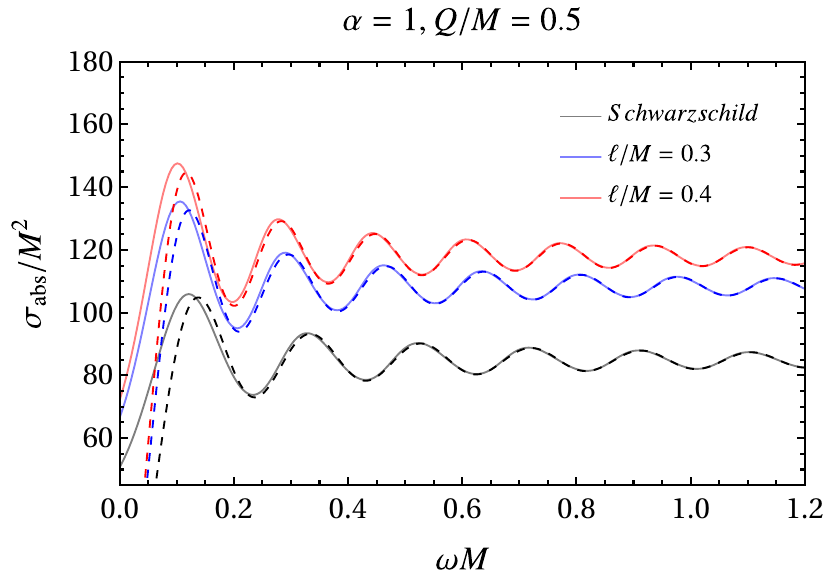}
	\vspace{0.2cm}
	\includegraphics[width=\columnwidth]{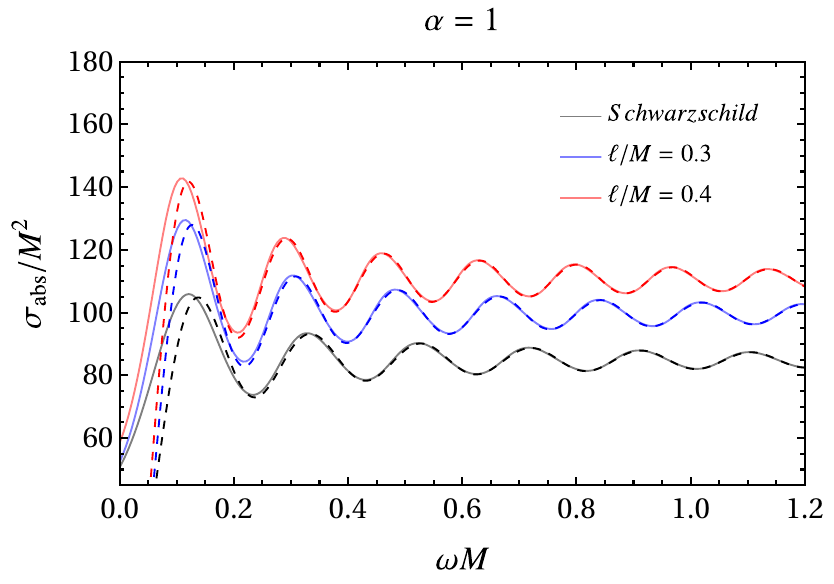}
	\vspace{0.2cm}
	\includegraphics[width=\columnwidth]{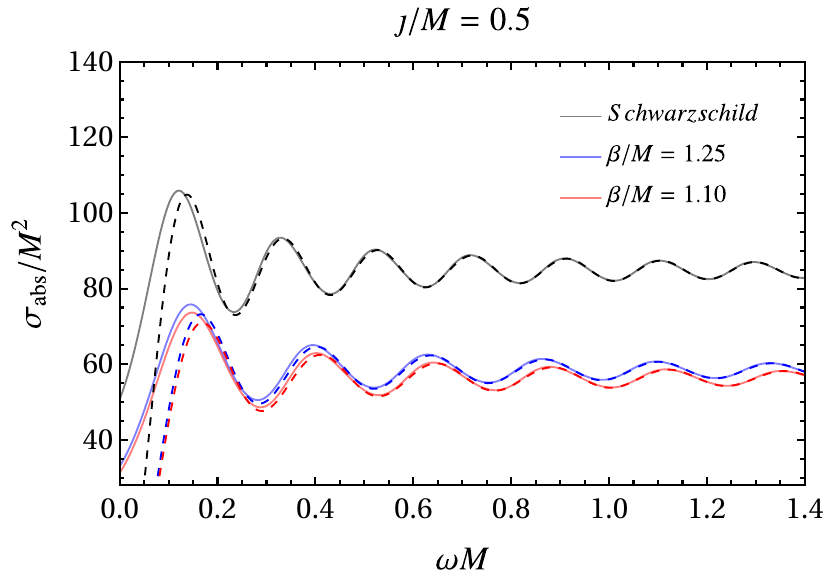}
	\caption{Total absorption cross section for DEC (top; $\alpha=1$ and $Q/M=0.5$, with varying $\ell$), SEC (middle; $\alpha=1$, with varying $\ell$), and WEC (bottom; $\jmath/M=0.5$, with varying $\beta/M$) solutions compared to the Schwarzschild case ($\alpha = 0$).}
	\label{fig:total-abs}
\end{figure}

In Fig.~\ref{fig:total-abs} we show the total absorption cross sections for some configurations of the three hairy BHs obtained via gravitational decoupling as a function of frequency and compare them with the scalar absorption cross section for the Schwarzschild BH with the same mass. The solid curves represent the numerical results, while the dashed curves, in corresponding colors, show the sinc approximation. As we can see, in the mid-to-high frequency regime, our numerical results are in very good agreement with the sinc approximation, which supports the accuracy of our numerical approach. 

\begin{figure}[ht!]
	\centering
	\includegraphics[width=\columnwidth]{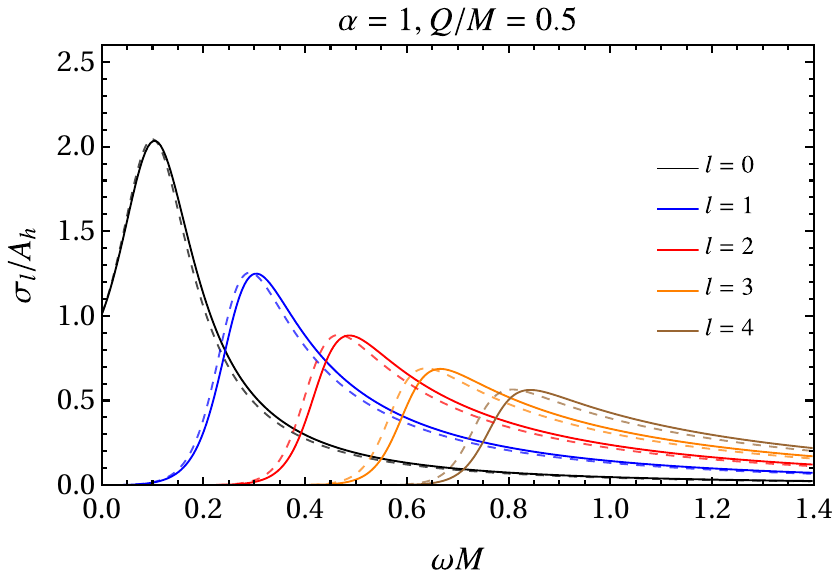}
	\vspace{0.2cm}
	\includegraphics[width=\columnwidth]{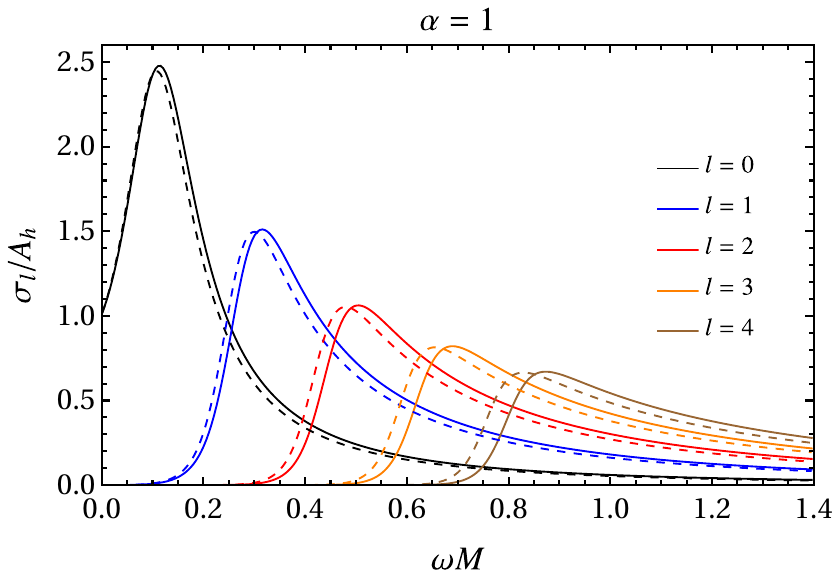}
	\vspace{0.2cm}
	\includegraphics[width=\columnwidth]{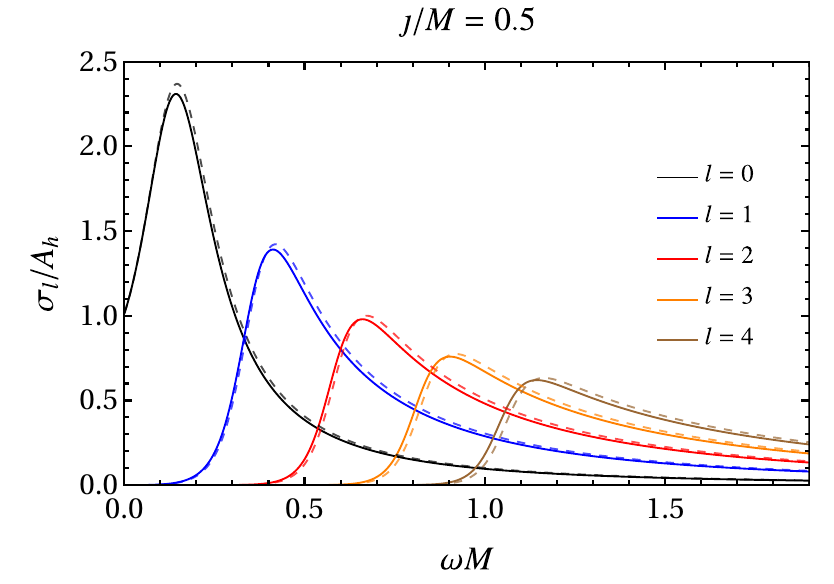}
	\caption{Partial absorption cross section for DEC (top; $\alpha = 1$ and $Q/M=0.5$), SEC (middle; $\alpha = 1$), and WEC (bottom; $\jmath/M = 0.5$) solutions. Solid lines correspond to $\ell/M = 0.3$, in the DEC and SEC cases, and to $\beta/M = 1.25$ in the WEC case; while dashed lines correspond to $\ell/M = 0.4$, in the DEC and SEC cases, and to $\beta/M = 1.10$ in the WEC case.}
	\label{fig:partial-abs}
\end{figure}

The top panel of Fig.~\ref{fig:total-abs} displays the total absorption cross section for the DEC solution with parameters $\alpha=1$ and $Q=0.5 M$ and two values of $\ell$, namely $\ell=0.3M$ and $\ell=0.4M$. We notice that larger values of the parameter $\ell$ result in a larger absorption cross section. This is consistent with the decrease of the scattering potential as $\ell$ increases, as shown in Fig.~\ref{fig:scalar-potentials}. The middle panel of Fig.~\ref{fig:total-abs} shows the results for the SEC solution with $\alpha=1$ and two values of $\ell$, namely $\ell=0.3M$ and $\ell=0.4M$. Again, increasing $\ell$ increases the absorption cross section. This is also consistent with the decrease in the scattering potential observed in Fig. \ref{fig:scalar-potentials} as $\ell$ grows. Finally, the bottom panel of Fig.~\ref{fig:total-abs} shows the results for the WEC solution with $\jmath=0.5M$ and two values of $\beta$, namely $\beta=1.1M$ and $\beta=1.25M$. This plot reveals that the increasing of the value of the parameter $\beta$ decreases the absorption cross section for the WEC solution. This behavior aligns with the increase in scattering potential as $\beta$ grows, as depicted in Fig.~\ref{fig:scalar-potentials}. In Fig.~\ref{fig:partial-abs} we plot the partial absorption cross sections. As we can see as $\omega M\rightarrow0$, our results show that $\sigma_{0}/A_h \rightarrow 1$, in agreement with the well-known low-frequency limit~\cite{Das1997,Higuchi2001, Higuchi2002}. 

\begin{figure}[ht!]
	\centering
	\includegraphics[width=\columnwidth]{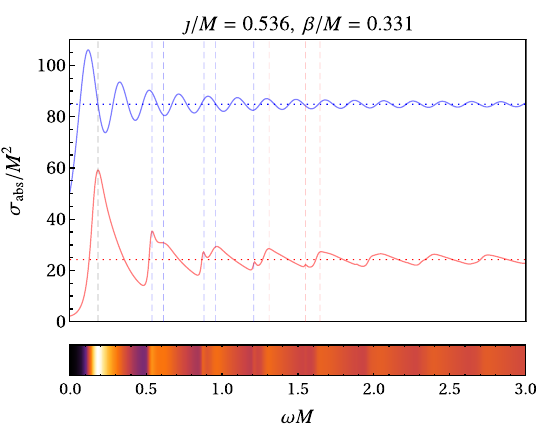}
	\caption{Total absorption cross section for a configuration ($\jmath/M=0.536$ and $\beta/M = 0.331$) that satisfies the WEC outside the event horizon and exhibits resonances, compared with the Schwarzschild solution. Dashed vertical lines indicate the real part of the trapped-mode frequencies, color coded as: $l=0$ (black), $l=1$ (blue), $l=2$ (red). Dotted horizontal lines correspond to the geometric capture cross section. Below the plot we show the corresponding absorption band.}
	\label{fig:abs-WEC-satisfied}
\end{figure}

\begin{figure}[ht!]
	\centering
	\includegraphics[width=\columnwidth]{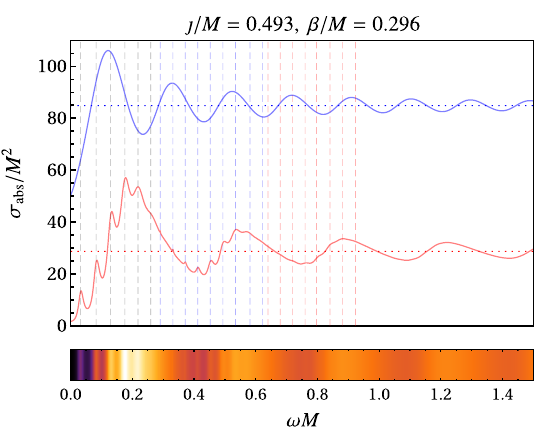}
	\caption{Total absorption cross section for a configuration ($\jmath/M=0.493$ and $\beta/M = 0.296$) that satisfies the WEC outside the event horizon and exhibits resonances, shown in comparison with the Schwarzschild solution. Dashed vertical lines indicate the real part of the trapped-mode frequencies, color coded as: $l=0$ (black), $l=1$ (blue), $l=2$ (red). Dotted horizontal lines correspond to the geometric capture cross section. Below the plot we show the corresponding absorption band.}
	\label{fig:abs-WEC-satisfied-lower}
\end{figure}

In Figs.~\ref{fig:abs-WEC-satisfied} and \ref{fig:abs-WEC-satisfied-lower}, we present the total absorption cross section of WEC configurations near the lower and upper bounds of the region shown in Fig.~\ref{fig:WEC-satisfied} and compare with that of the Schwarzschild solution with the same mass. In both cases, the effective potential of these configurations exhibit a well, leading to the presence of quasibound states. We list some of the quasibound state frequencies, obtained via direct integration, in Table~\hyperref[tab:frequencies-WEC]{I} for the configuration with $\jmath/M = 0.493$ and $ \beta/M = 0.296$, confirming the existence of long-lived modes, characterized by extremely small imaginary parts. The selected frequencies shown in Table~\hyperref[tab:frequencies-WEC]{I} are associated to the clearly visible spectral lines in the absorption profile. The agreement between the observed spectral lines and the computed quasibound state frequencies validates the consistency of our numerical results. In Figs.~\ref{fig:abs-WEC-satisfied} and \ref{fig:abs-WEC-satisfied-lower}, vertical dashed lines correspond to the real part of the trapped modes that lead to observable imprints on the absorption spectrum. Additional modes can be found, but are not shown, since they correspond to very small peaks in the transmission coefficients, which are further suppressed in the total absorption cross section. We notice that, as the frequency increases the spectral lines are less evident in the absorption profile, in contrast with the low-frequency regime where a large number of spectral lines can be observed.  It is worth noting that the width and number of spectral lines in the absorption spectrum is related to the depth of the potential well (cf. Fig.~\ref{fig:potential-eco}), since a deeper well can trap waves more effectively. 

\begin{figure}[h!] 
	\centering
	\includegraphics[width=\columnwidth]{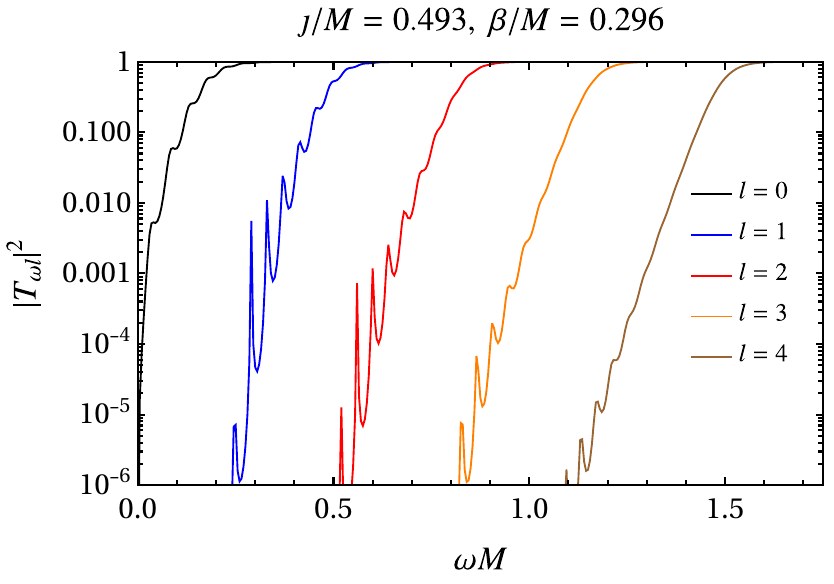}
	\caption{Transmission coefficients of a WEC BH configuration, with $\jmath/M = 0.493$ and $\beta/M = 0.296$; presenting quasibound states. }
	\label{fig:trans-WEC-ECO}
\end{figure}

Finally, in Fig.~\ref{fig:trans-WEC-ECO} we show the transmission coefficients of the WEC configuration with parameters $\jmath/M=0.493$ and $\beta/M=0.296$. As we can see, due to the existence of the long-lived modes, resonant peaks are present in the transmission coefficients, which results in the presence of spectral lines in the absorption cross section. 

\begin{table}[htbp]
	\centering
	\caption{Trapped modes associated with the observable spectral lines in the absorption profile of the WEC solution
		($\jmath/M = 0.493$ and $\beta/M = 0.296$).}
	\label{tab:frequencies-WEC}
	
	\begin{tabular}{|c|c|c|}
		\hline
		$l$ & $M\omega_{R}$ & $-iM\omega_{I}$ \\
		\hline
		\multirow{6}{4em}{\hspace{0.6cm}0} 
		& \hspace{0.2cm}$3.187\times10^{-2}$\hspace{0.2cm} & \hspace{0.2cm}$1.002\times10^{-2}$\hspace{0.2cm} \\ 
		& \hspace{0.2cm}$8.215\times10^{-2}$\hspace{0.2cm} & \hspace{0.2cm}$1.314\times10^{-2}$\hspace{0.2cm} \\ 
		& \hspace{0.2cm}$1.291\times10^{-1}$\hspace{0.2cm} & \hspace{0.2cm}$1.661\times10^{-2}$\hspace{0.2cm} \\ 
		& \hspace{0.2cm}$1.751\times10^{-1}$\hspace{0.2cm} & \hspace{0.2cm}$2.089\times10^{-2}$\hspace{0.2cm} \\
		& \hspace{0.2cm}$2.177\times10^{-1}$\hspace{0.2cm} & \hspace{0.2cm}$2.299\times10^{-2}$\hspace{0.2cm} \\  
		& \hspace{0.2cm}$2.589\times10^{-1}$\hspace{0.2cm} & \hspace{0.2cm}$2.436\times10^{-2}$\hspace{0.2cm} \\ 
		\hline
		\multirow{9}{4em}{\hspace{0.6cm}1}
		& \hspace{0.2cm}$2.897\times10^{-1}$\hspace{0.2cm} & \hspace{0.2cm}$4.523\times10^{-4}$\hspace{0.2cm} \\ 
		& \hspace{0.2cm}$3.307\times10^{-1}$\hspace{0.2cm} & \hspace{0.2cm}$1.746\times10^{-3}$\hspace{0.2cm} \\ 
		& \hspace{0.2cm}$3.710\times10^{-1}$\hspace{0.2cm} & \hspace{0.2cm}$4.584\times10^{-3}$\hspace{0.2cm} \\ 
		& \hspace{0.2cm}$4.113\times10^{-1}$\hspace{0.2cm} & \hspace{0.2cm}$8.632\times10^{-3}$\hspace{0.2cm} \\ 
		& \hspace{0.2cm}$4.517\times10^{-1}$\hspace{0.2cm} & \hspace{0.2cm}$1.326\times10^{-2}$\hspace{0.2cm} \\ 
		& \hspace{0.2cm}$4.928\times10^{-1}$\hspace{0.2cm} & \hspace{0.2cm}$1.705\times10^{-2}$\hspace{0.2cm} \\ 
		& \hspace{0.2cm}$5.335\times10^{-1}$\hspace{0.2cm} & \hspace{0.2cm}$1.899\times10^{-2}$\hspace{0.2cm} \\ 
		& \hspace{0.2cm}$5.804\times10^{-1}$\hspace{0.2cm} & \hspace{0.2cm}$2.012\times10^{-2}$\hspace{0.2cm} \\ 
		& \hspace{0.2cm}$6.216\times10^{-1}$\hspace{0.2cm} & \hspace{0.2cm}$2.153\times10^{-2}$\hspace{0.2cm} \\ 
		\hline
		\multirow{8}{4em}{\hspace{0.6cm}2}
		& \hspace{0.2cm}$6.390\times10^{-1}$\hspace{0.2cm} & \hspace{0.2cm}$5.092\times10^{-3}$\hspace{0.2cm} \\ 
		& \hspace{0.2cm}$6.787\times10^{-1}$\hspace{0.2cm} & \hspace{0.2cm}$9.124\times10^{-3}$\hspace{0.2cm} \\ 
		& \hspace{0.2cm}$7.188\times10^{-1}$\hspace{0.2cm} & \hspace{0.2cm}$1.352\times10^{-2}$\hspace{0.2cm} \\ 
		& \hspace{0.2cm}$7.593\times10^{-1}$\hspace{0.2cm} & \hspace{0.2cm}$1.765\times10^{-2}$\hspace{0.2cm} \\ 
		& \hspace{0.2cm}$7.963\times10^{-1}$\hspace{0.2cm} & \hspace{0.2cm}$1.679\times10^{-2}$\hspace{0.2cm} \\ 
		& \hspace{0.2cm}$8.399\times10^{-1}$\hspace{0.2cm} & \hspace{0.2cm}$2.330\times10^{-2}$\hspace{0.2cm} \\ 
		& \hspace{0.2cm}$8.809\times10^{-1}$\hspace{0.2cm} & \hspace{0.2cm}$2.466\times10^{-2}$\hspace{0.2cm} \\ 
		& \hspace{0.2cm}$9.226\times10^{-1}$\hspace{0.2cm} & \hspace{0.2cm}$2.519\times10^{-2}$\hspace{0.2cm} \\ 
		\hline
	\end{tabular}
\end{table}

\section{Final Remarks}
\label{sec:remarks}
We have studied the propagation of massless scalar fields in BH spacetimes generated through gravitational decoupling, a method that exploits a quasi-linear behavior of Einstein's equations to derive new solutions from known seed solutions. We investigated three solutions obtained after imposing the SEC, DEC and WEC, yielding BHs with primary hair. The presence of transcendental functions (exponential and logarithmic) in their metric components leads to distinctive features. 

An analysis of massless particle trajectories in those backgrounds revealed a key characteristic: stable light rings may be present in those hairy BH solutions. Using numerical methods, we scanned the parameter space for configurations exhibiting stable light rings outside the event horizon. All such configurations, however, violate the SEC within a finite region beyond the outer horizon.

By numerically integrating the radial equation of the scalar field, we computed the partial and total absorption cross sections via the partial-wave method. Our results show that for all solutions, the total absorption cross section converges to the event horizon's surface area at low frequencies and oscillates around the geometric absorption cross section in the high-frequency regime. We further examined the influence of primary hair on the transmission coefficients.

Although the hair modifies the ADM mass in two cases (DEC and SEC), this alone cannot explain our findings. The additional parameters associated to the primary hair fundamentally alters the spacetime geometry, which manifests most strikingly in the emergence of novel absorption features beyond what mass modification alone would predict. Crucially, the hair enables potential wells outside the event horizon, which are absent in the seed solutions. We show that these hair-induced potential wells allow for quasibound states to emerge, manifesting as spectral lines in absorption spectra, associated to Breit-Wigner-like resonances in transmission coefficients. By employing the direct integration method, we obtained the imaginary part of the trapped modes, confirming that they are extremely small.

Our results indicate that, within standard GR, hairy BH geometries can produce observational signatures characteristic of wormholes and other horizonless ultracompact objects. Scattering processes in these spacetimes provide a powerful framework for probing how non-trivial hair parameters modify the spacetime structure. Consequently, future BH spectroscopy detections and refined scattering analyses emerge as decisive tools to test the no-hair paradigm.
\section{acknowledgements}

The authors would like to acknowledge Funda\c{c}\~ao Amaz\^onia de Amparo a Estudos e Pesquisas (FAPESPA),  Conselho Nacional de Desenvolvimento Cient\'ifico e Tecnol\'ogico (CNPq) and Coordena\c{c}\~ao de Aperfei\c{c}oamento de Pessoal de N\'{\i}vel Superior (Capes) - Finance Code 001, in Brazil, for partial financial support. This research has further been supported by the European Horizon Europe staff exchange (SE) programme HORIZON-MSCA-2021-SE-01 Grant No. NewFunFiCO-101086251. L.C. would like to thank the University of Aveiro, in Portugal, for the kind hospitality during the completion of this work.
R.B.M. is supported by CNPq/PDJ 151250/2024-3.

\bibliographystyle{report}
\bibliography{export.bib}

\end{document}